\pdfoutput=1
\documentclass[aps,prd,amsmath,floats,floatfix, twocolumn,
superscriptaddress,nofootinbib,showpacs,longbibliography]{revtex4-1}
\usepackage[T1]{fontenc}
\usepackage[utf8]{inputenc}
\usepackage{txfonts}
\usepackage{verbatim}
\usepackage[dvipsnames, usenames, table]{xcolor}
\definecolor{linkcolor}{rgb}{0.0,0.3,0.5}
\usepackage[hypertexnames=false, unicode, colorlinks=true, linkcolor=linkcolor,
citecolor=linkcolor, filecolor=linkcolor,urlcolor=linkcolor,
pdfusetitle]{hyperref}
\usepackage{orcidlink}
\usepackage[all]{hypcap}
\usepackage{graphicx}
\usepackage{xspace}
\usepackage{amssymb}
\usepackage{microtype}
\usepackage{subfigure}

\usepackage{url}

\usepackage[english]{babel}
\usepackage{blindtext}

\usepackage[normalem]{ulem} %for \sout
\usepackage{bm} % boldmath
\usepackage{mathrsfs}
\usepackage{xspace} % for \xspace
\usepackage{fontawesome}

\DeclareMathAlphabet{\mathpzc}{OT1}{pzc}{m}{it}
\usepackage[nomessages]{fp}

\newcommand{\sk}[1]{}

\newcommand{\orcid}[1]{\href{https://orcid.org/#1}{\includegraphics[width=10pt]{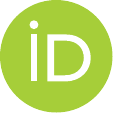}}}
\begin{document}
\title{GW190711\_030756 and GW200114\_020818: astrophysical interpretation of two asymmetric binary black hole mergers in the IAS catalog}

\author{Tousif Islam\,\orcid{0000-0002-3434-0084}}
\email{tousifislam@ucsb.edu}
\affiliation{Kavli Institute for Theoretical Physics,University of California Santa Barbara, Kohn Hall, Lagoon Rd, Santa Barbara, CA 93106}

\author{Tejaswi Venumadhav\,\orcid{0000-0002-1661-2138}}
\affiliation{\mbox{Department of Physics, University of California at Santa Barbara, Santa Barbara, CA 93106, USA}}
\affiliation{\mbox{International Centre for Theoretical Sciences, Tata Institute of Fundamental Research, Bangalore 560089, India}}

\author{Digvijay Wadekar\,\orcid{0000-0002-2544-7533}}
\affiliation{\mbox{Center for Gravitational Physics, University of Texas at Austin, Austin, TX 78712, USA}}
\affiliation{\mbox{Department of Physics and Astronomy, Johns Hopkins University, 3400 N. Charles Street, Baltimore, Maryland, 21218, USA}}

\author{Ajit Kumar Mehta\,\orcid{0000-0002-7351-6724}}
\affiliation{\mbox{Chennai Mathematical Institute, Siruseri 603013, Chennai, India}}

\author{Javier Roulet\,\orcid{0000-0003-3268-4796}}
\affiliation{\mbox{Kavli Institute for Cosmological Physics, The University of Chicago, 5640 South Ellis Avenue, Chicago, Illinois 60637, USA}}
\affiliation{\mbox{School of Natural Sciences, Institute for Advanced Study, 1 Einstein Drive, Princeton, NJ 08540, USA}}

\author{Jonathan Mushkin\,\orcid{0009-0009-9057-9581}}
\affiliation{\mbox{Department of Particle Physics \& Astrophysics, Weizmann Institute of Science, Rehovot 76100, Israel}}

\author{Mark Ho-Yeuk Cheung\,\orcid{0000-0002-7767-3428}}
\affiliation{\mbox{School of Natural Sciences, Institute for Advanced Study, 1 Einstein Drive, Princeton, NJ 08540, USA}}

\author{Barak Zackay\,\orcid{0000-0001-5162-9501}}
\affiliation{\mbox{Department of Particle Physics \& Astrophysics, Weizmann Institute of Science, Rehovot 76100, Israel}}

\author{Matias Zaldarriaga\,\orcid{0009-0007-8315-6703}}
\affiliation{\mbox{School of Natural Sciences, Institute for Advanced Study, 1 Einstein Drive, Princeton, NJ 08540, USA}}

% Because hyperref only gets the *last* author, we need to be explicit.
\hypersetup{pdfauthor={Islam et al.}}

\date{\today}

%==========================================================================
\begin{abstract}
We provide a comprehensive analysis of GW190711\_030756 and GW200114\_020818, two of the most significant binary black hole merger candidates in the IAS catalog, with probabilities of astrophysical origin $p_{\rm astro}=0.99$ and $0.71$, respectively, and signal-to-noise ratios of approximately $10.0$ and $13.4$. 
We employ numerical relativity surrogate models to infer both the source properties and the remnant properties of these two candidates. 
We find that both GW190711\_030756 and GW200114\_020818 are asymmetric-mass binaries, with inferred mass ratios of $0.35^{+0.32}_{-0.15}$ and $\leq 0.20$. 
In addition, GW200114\_020818 is inferred to have a source-frame total mass of approximately $220M_{\odot}$ and highly spinning black holes, with primary (secondary) dimensionless spin magnitudes of $0.96^{+0.03}_{-0.07}$ ($0.84^{+0.13}_{-0.34}$), closely resembling GW231123\_135430. 
We further find that GW200114\_020818 has a confidently negative effective inspiral spin of $\chi_{\rm eff}=-0.60^{+0.22}_{-0.13}$ and exhibits strong spin precession, characterized by an effective precession parameter of $\chi_{\rm p}=0.60^{+0.21}_{-0.19}$. 
GW200114\_020818 (when considered alongside GW231123\_135430) points towards an emerging population of massive, rapidly spinning BBH mergers.
While GW231123\_135430 is consistent with mergers in globular clusters, producing systems like GW200114\_020818 in such environments remains difficult even under hierarchical merger scenarios. The probability that the remnant black hole of GW190711\_030756 (GW200114\_020818) is retained in its host environment is $0.079$ ($0.0002$), $0.62$ ($0.965$), and $0.997$ ($1$) if the merger occurred in a globular cluster, a nuclear star cluster, or an elliptical galaxy, respectively.
\end{abstract}

\maketitle

%%%%%%%%%%%%%%%%%%%%%%%%%%%%%%%%%%%%%%%%%%%%%%%%%%%%%%%%%%%%%%%%%%%%%%%%%%%%%%%%%%%
%%%%%%%%%%%%%%%%%%%%%%%%%%%%%%%%%%%%%%%%%%%%%%%%%%%%%%%%%%%%%%%%%%%%%%%%%%%%%%%%%%%
\section{Introduction}
\label{sec:intro}
%%%%%%%%%%%%%%%%%%%%%%%%%%%%%%%%%%%%%%%%%%%%%%%%%%%%%%%%%%%%%%%%%%%%%%%%%%%%%%%%%%%
%%%%%%%%%%%%%%%%%%%%%%%%%%%%%%%%%%%%%%%%%%%%%%%%%%%%%%%%%%%%%%%%%%%%%%%%%%%%%%%%%%%
Gravitational-wave (GW) astronomy has entered a data-rich era, with the LIGO--Virgo--KAGRA (LVK) Collaboration~\cite{Harry:2010zz,VIRGO:2014yos,KAGRA:2020tym} having detected around $200$ compact binary merger events to date~\cite{LIGOScientific:2018mvr,LIGOScientific:2020ibl,LIGOScientific:2021usb,KAGRA:2021vkt,LIGOScientific:2025slb}. In addition, independent research groups employing different detection strategies, often targeting the low signal-to-noise ratio (SNR) regime, have reported several tens of additional GW candidates~\cite{Zackay:2019tzo,Venumadhav:2019lyq,Olsen:2022pin,Mehta:2023zlk,Wadekar:2023gea,Nitz:2018imz,Nitz:2020oeq,Nitz:2021uxj,Nitz:2021zwj,Koloniari:2024kww}. The majority of both confirmed events and candidates correspond to binary black hole (BBH) mergers. 
These events are subsequently analyzed using a combination of time-domain and frequency-domain waveform models that are either trained directly on numerical relativity (NR) data, such as NR surrogate models~\cite{Field:2011mf,Field:2013cfa,Varma:2018aht,Varma:2018mmi,Islam:2021mha,Islam:2022laz,Blackman:2015pia,Blackman:2017pcm,Blackman:2017dfb,Yoo:2022erv}, or calibrated to NR simulations (such as for phenomenological models~\cite{Hannam:2013oca,Khan:2018fmp,Schmidt:2014iyl,Estelles:2021gvs,Estelles:2020twz,Pratten:2020ceb,Yu:2023lml,Thompson:2023ase,Hamilton:2025xru,Khan:2019kot} and effective-one-body (EOB) models~\cite{Taracchini:2013rva,Bohe:2016gbl,Cotesta:2018fcv,Nagar:2019wds,Nagar:2020pcj,Nagar:2021gss,Pan:2013rra,Babak:2016tgq,Ossokine:2020kjp,Ramos-Buades:2023ehm}).

A subset of these events and candidates is of particular astrophysical interest, as they appear to be outliers relative to the bulk of the population.
For example, GW190412\_053044~\cite{LIGOScientific:2020stg,Islam:2020reh} and GW190814\_211039~\cite{LIGOScientific:2020zkf,Yoo:2022erv} are characterized by highly asymmetric mass ratios, with $q := m_2/m_1$ (where $m_1$ and $m_2$ denote the primary and secondary black hole masses, respectively) of approximately $0.25$ and $0.1$. Interestingly, GW191219\_163120 has been reported to have an extreme mass ratio of $q \sim 0.035$, although the possibility that this signal is a noise artifact remains under discussion~\cite{KAGRA:2021vkt}. GW191109\_010717~\cite{KAGRA:2021vkt,Islam:2023zzj} exhibits evidence for a negative effective spin, while GW191129\_134042~\cite{KAGRA:2021vkt} marks the least massive BBH merger observed to date, with a total mass of approximately $17\,M_{\odot}$.
Several events, including GW190412\_053044~\cite{LIGOScientific:2020stg,Islam:2020reh}, GW190521\_030229~\cite{LIGOScientific:2020iuh}, GW191109\_010717~\cite{KAGRA:2021vkt,Islam:2023zzj}, and GW200129\_065458~\cite{KAGRA:2021vkt,Islam:2023zzj}, have shown signatures consistent with spin-induced precession, and a large recoil kick has been inferred for GW200129\_065458~\cite{Varma:2022pld,Islam:2023zzj}. Moreover, GW190521\_030229~\cite{LIGOScientific:2020iuh} and GW231123\_135430~\cite{LIGOScientific:2025rsn} provide evidence for the existence of intermediate-mass black holes, with inferred total masses of approximately $150\,M_{\odot}$ and $230\,M_{\odot}$, respectively. The event GW231123\_135430~\cite{LIGOScientific:2025rsn} is also notable as an outlier due to both black holes being inferred to be highly spinning.
More recent candidates, such as GW241011\_065126 and GW241110\_043853~\cite{LIGOScientific:2025brd}, have been hypothesized to form through hierarchical merger scenarios, with GW241011\_065126 also inferred to possess negative spins, similar to GW191109\_010717~\cite{KAGRA:2021vkt,Islam:2023zzj}. The current GW catalog further includes GW250114\_082203~\cite{LIGOScientific:2025wao}, an event with an unusually high SNR of approximately $76$, enabling some of the most stringent tests to date of Hawking's area law and the Kerr nature of astrophysical black holes.

Furthermore, most analyses to date have used waveform models that do not include the effects of orbital eccentricity. As eccentric waveform models continue to be developed and incorporated into source characterization, a subset of events has been reported to exhibit signatures of residual eccentricity~\cite{Romero-Shaw:2020thy,Gayathri:2020coq,Gamba:2021gap,Ramos-Buades:2023yhy,Gupte:2024jfe,Morras:2025xfu,Jan:2025zcm}. There is also interest in understanding whether the unusual source properties of GW231123\_135430~\cite{LIGOScientific:2025rsn} could be explained by gravitational lensing due to intervening galaxies or galaxy clusters~\cite{Chan:2025kyu}.

%%%%%%%%%%%%%%%%%%%%%%%%%%%%%%%%%%%%%%%%%%%%%%%%%%%%%%%%%%%%%%%%%%%%%%%%%%%%%%%%%%%
\begin{figure*}
    \centering
    \includegraphics[width=0.95\textwidth]{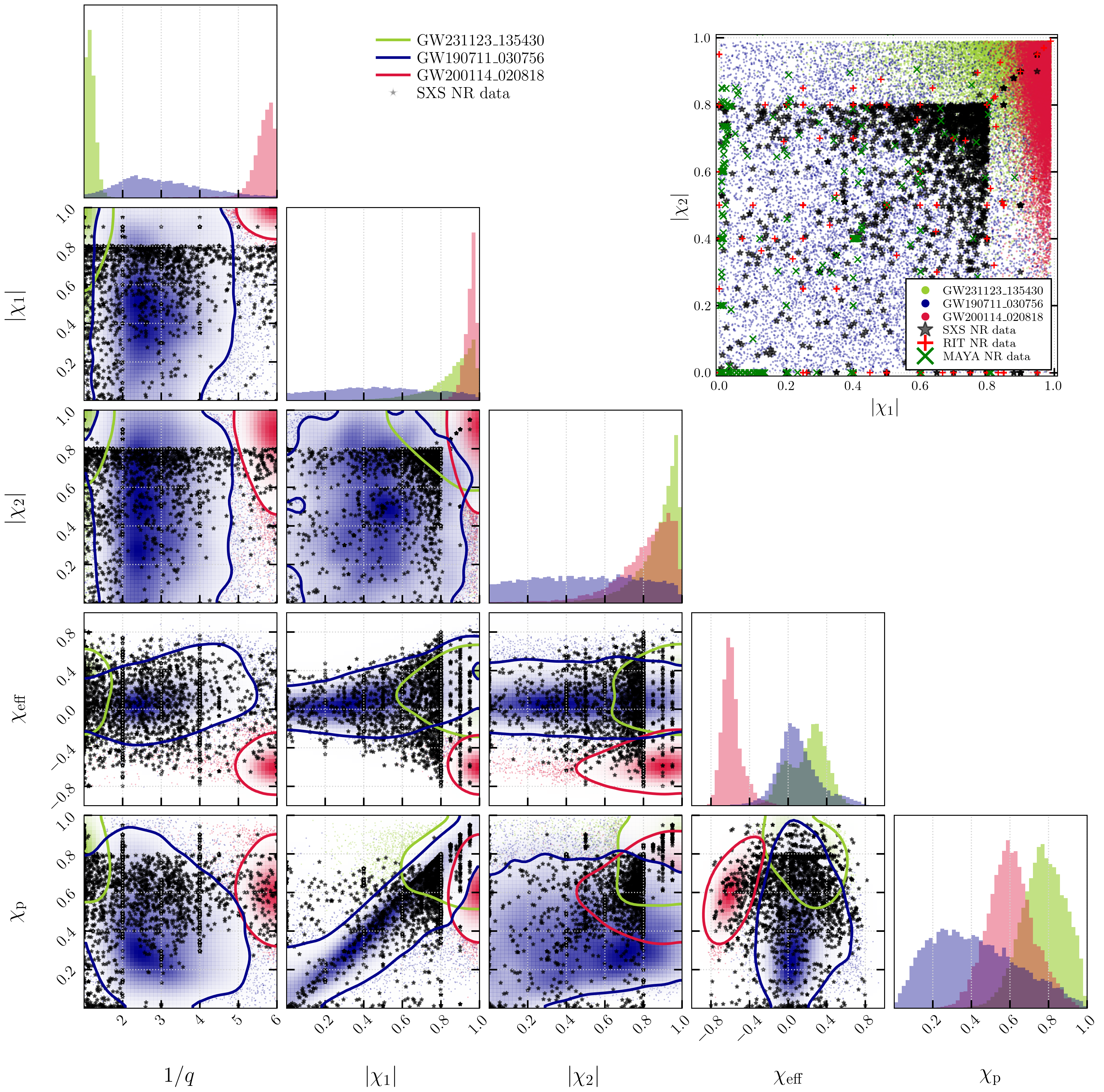}
    \caption{We show the posteriors for five key intrinsic source properties (along with the recovered log likelihood) of two asymmetric-mass BBH events, GW190711\_030756 (dark blue) and GW200114\_020818 (crimson), inferred using the numerical-relativity precessing surrogate waveform model \texttt{NRSur7dq4}. For comparison, we also show the posteriors for GW231123\_135430 (yellowgreen), together with the available SXS NR simulations within $q \in [0.1667,1]$. In addition, we show the spin magnitudes of the available NR simulations from three different NR catalogs (see inset). GW190711\_030756 was first reported in the $(2,2)$–only IAS search and subsequently increased in significance in the IAS-HM search (including quasi-circular higher-order modes) with astrophysical probability $p_{\mathrm{astro}}> 0.99$. It has also subsequently been confirmed by other pipelines \cite{Kum24, Koloniari:2024kww, Mis24_CWB}. GW200114\_020818 was first reported in a targeted IMBH search by the LVK collaboration and later detected by the IAS-HM search over the full BBH mass range with $p_{\mathrm{astro}}= 0.71$. These events therefore provide strong motivation to extend the coverage of NR simulations in the highly spinning and asymmetric mass ratio spaces. More details are in Section~\ref{sec:nrsurpe}.}
    \label{fig:combined_posteriors}
\end{figure*}
%%%%%%%%%%%%%%%%%%%%%%%%%%%%%%%%%%%%%%%%%%%%%%%%%%%%%%%%%%%%%%%%%%%%%%%%%%%%%%%%%%%

In this context, we take a closer look at the GW candidates reported by the IAS detection pipeline~\cite{Zackay:2019tzo,Venumadhav:2019lyq,Olsen:2022pin,Mehta:2023zlk,Wadekar:2023gea,Wadekar:2024zdq}, which is specifically tuned to recover additional subthreshold, low-SNR BBH mergers, while also identifying high-SNR events reported by LVK. The IAS pipelines additionally incorporate higher-order spherical harmonic modes that are currently neglected in the current LVK detection pipelines. We identify the two most significant candidates in the IAS catalog~\cite{Zackay:2019tzo,Venumadhav:2019lyq,Olsen:2022pin,Mehta:2023zlk,Wadekar:2023gea}, namely GW190711\_030756 and GW200114\_020818. These are the loudest candidates with SNRs of approximately $10$ and $13$, respectively.
The event GW190711\_030756 is inferred to have a probability of astrophysical origin of $p_{\rm astro} = 0.97$ in the IAS $(2,2)$-mode-only search pipeline~\cite{Olsen:2022pin} and $p_{\rm astro} = 0.99$ in the IAS higher-order-modes pipeline~\cite{Wadekar:2023gea}, which includes the $(\ell, |m|) = (3,3)$ and $(4,4)$ modes in addition to the dominant $(2,2)$ mode. GW190711\_030756 also has subsequently been confirmed by other pipelines \cite{Kum24, Koloniari:2024kww, Mis24_CWB}.

On the other hand, GW200114\_020818 is identified with $p_{\rm astro} = 0.71$ in the IAS higher-order-modes detection pipeline~\cite{Wadekar:2023gea}. The inverse false-alarm rates (IFARs) for GW190711\_030756 and GW200114\_020818 in the IAS higher-order-modes pipeline are $10.96$ and $0.16$ years, respectively~\cite{Olsen:2022pin,Mehta:2023zlk,Wadekar:2023gea}. Note, however, that the IFAR and $p_\mathrm{astro}$ values depend on the astrophysical prior used in the IAS search, especially for GW200114\_020818, as it lies in the high-mass regime where the astrophysical models are currently poorly constrained.
Furthermore, GW200114\_020818 was first identified as a candidate in the targeted LVK intermediate-mass black hole search, with an IFAR of $7.6$ years~\cite{LIGOScientific:2021tfm}.

Initial IAS analyses~\cite{Olsen:2022pin,Mehta:2023zlk,Wadekar:2023gea} of GW190711\_030756 and GW200114\_020818 employed \texttt{IMRPhenomXPHM}~\cite{Pratten:2020ceb} and \texttt{IMRPhenomXODE}~\cite{Yu:2023lml}, two frequency-domain phenomenological waveform models, which inferred the systems to be asymmetric, high-mass binaries\footnote{\href{https://github.com/seth-olsen/new\_BBH\_mergers\_O3a\_IAS\_pipeline}{https://github.com/seth-olsen/new\_BBH\_mergers\_O3a\_IAS\_pipeline}}
\footnote{\href{https://github.com/JayWadekar/GW\_higher\_harmonics\_search}{https://github.com/JayWadekar/GW\_higher\_harmonics\_search}}. Frequency-domain phenomenological models rely on simplified modeling approximations in the merger and ringdown regime~\cite{Pratten:2020ceb,Yu:2023lml,MacUilliam:2024oif}, which dominates the strain signal for GW190711\_030756 and GW200114\_020818. Moreover, for asymmetric-mass binaries, these models have been shown to exhibit noticeable differences when compared against NR, which is typically taken as the ground truth~\cite{Pratten:2020ceb,Yu:2023lml,MacUilliam:2024oif}. It is therefore more appropriate to employ time-domain waveform approximants, such as \texttt{IMRPhenomTPHM}~\cite{Estelles:2021gvs}, \texttt{SEOBNRv4PHM}~\cite{Ossokine:2020kjp} (and its successor \texttt{SEOBNRv5PHM}~\cite{Ramos-Buades:2023ehm}), and \texttt{NRSur7dq4}~\cite{Varma:2019csw}, which provide improved accuracy in the merger and ringdown portions of the waveform. In particular, \texttt{NRSur7dq4} is considered as the most accurate waveform model for precessing-spin binaries within its training range, corresponding to mass ratios $0.16 \leq q \leq 1$ and detector-frame total masses of $M_{\rm tot}^{\rm det}:=m_1^{\rm det}+m_2^{\rm det} \ge 60\,M_{\odot}$~\cite{Varma:2019csw}.

% \begin{figure*}
%     \centering
%     \includegraphics[width=\textwidth]{combined_posteriors.png}
%     \caption{We show the posteriors for seven key source properties (along with the recovered log likelihood) of two asymmetric-mass BBH events, GW190711\_030756 (dark blue) and GW200114\_020818 (crimson), inferred using the numerical-relativity precessing surrogate waveform model \texttt{NRSur7dq4}.  GW190711\_030756 was first reported in the $(2,2)$–only IAS search and subsequently increased in significance in the IAS-HM search (including quasi-circular higher-order modes) with astrophysical probability $p_{\mathrm{astro}}> 0.99$. GW200114\_020818 was first reported in a targeted IMBH search by the LVK collaboration and later detected by the IAS-HM search over the full BBH mass range with $p_{\mathrm{astro}}= 0.71$. More details are in Section~\ref{sec:nrsurpe}. 
%     %\javier{Is $\log \mathcal L = \langle d \mid h \rangle - \frac 12 \langle h \mid h \rangle$? Then the numbers don't make sense to me given the SNRs.}
%     }
%     \label{fig:combined_posteriors}
% \end{figure*}

\begin{figure*}
    \centering
    \includegraphics[width=0.49\textwidth]{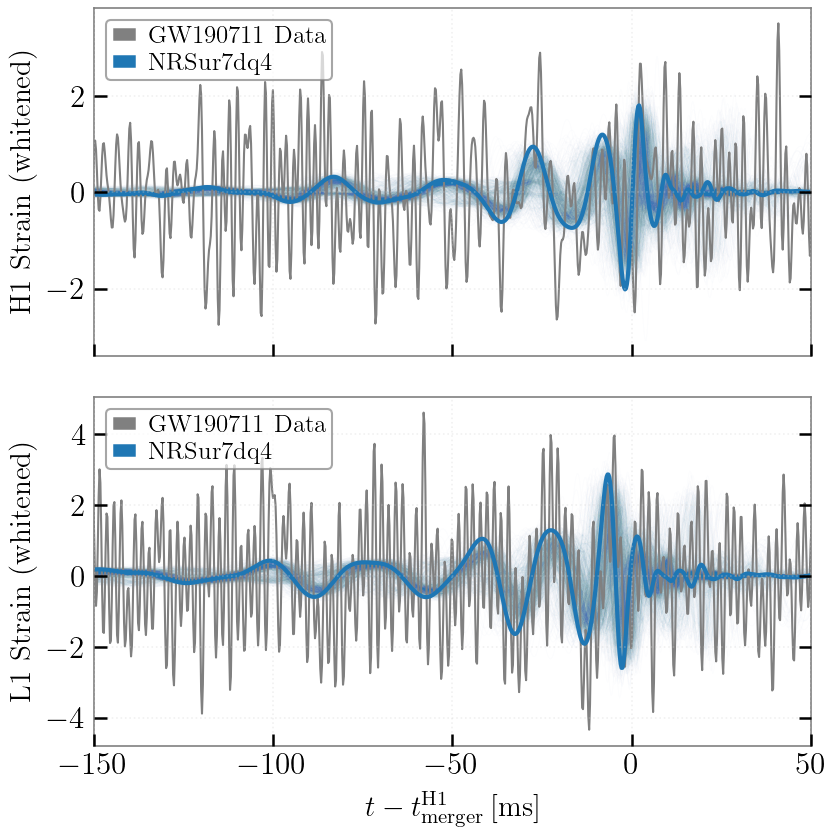}
    \includegraphics[width=0.49\textwidth]{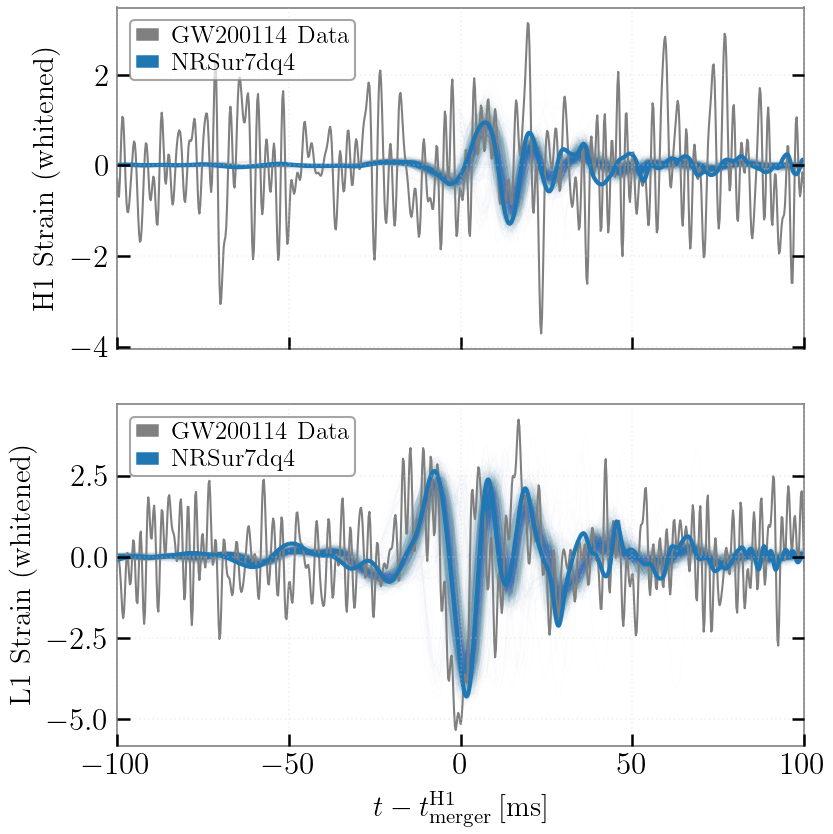}
    \caption{We show the whitened strain data (gray) for two asymmetric-mass BBH events, GW190711\_030756 (left) and GW200114\_020818 (right), as observed in the H1 and L1 detectors. We also overlay the best-fit template obtained from the numerical-relativity precessing surrogate waveform model \texttt{NRSur7dq4} (thick blue line). In addition, we display 1000 templates corresponding to 1000 randomly selected source parameters drawn from the inferred \texttt{NRSur7dq4} posterior. For details of the IAS detection pipelines that have found these two events, we refer the readers to Refs.~\cite{Olsen:2022pin,Mehta:2023zlk,Wadekar:2023gea}. More details are in Section~\ref{sec:nrsurpe}.}
    \label{fig:whitened_strain}
\end{figure*}

\begin{table*}[htbp]
\centering
\setlength{\tabcolsep}{3pt}

\caption{\textbf{Top}: Summary of parameter estimation results for GW190711\_030756. \textbf{Bottom}: Summary of parameter estimation results for GW200114\_020818. We show the inferred median and 90\% credible intervals for mass ratio $q$, source-frame total mass $M_{\rm tot}^{\rm source}$ (in $M_{\odot}$), source-frame individual masses $m_{1,2}^{\rm source}$ (in $M_{\odot}$), dimensionless spin magnitudes $|\chi_{1,2}|$, effective inspiral spin parameter $\chi_{\rm eff}$, effective transverse precession spin parameter $\chi_{\rm p}$, luminosity distance $D_{\rm L}$ (in Mpc), and log likelihood $\log_\mathcal{L}$.}

\begin{tabular}{lcccccccccc}

\hline\hline
Model & $q$ & $M_{\rm tot}^{\rm source}$ & $m_1^{\rm source}$ & $m_2^{\rm source}$ & $|\chi_1|$ & $|\chi_2|$ & $\chi_{\rm eff}$ & $\chi_{\rm p}$ & $D_{\rm L}$ & $\log\mathcal{L}$\\
\hline
\rowcolor{gray!20} NRSur7dq4 & $0.35^{+0.32}_{-0.15}$ & $85.43^{+35.71}_{-17.37}$ & $62.82^{+35.40}_{-19.51}$ & $22.44^{+9.49}_{-6.65}$ & $0.48^{+0.44}_{-0.41}$ & $0.49^{+0.43}_{-0.43}$ & $0.08^{+0.45}_{-0.28}$ & $0.39^{+0.43}_{-0.29}$ & $2322.69^{+2170.79}_{-1070.08}$ & $36.42^{+3.08}_{-4.80}$ \\
IMRPhenomXPHM & $0.34^{+0.39}_{-0.23}$ & $83.47^{+48.45}_{-18.47}$ & $61.54^{+53.00}_{-20.66}$ & $20.82^{+12.34}_{-9.10}$ & $0.34^{+0.45}_{-0.30}$ & $0.49^{+0.45}_{-0.44}$ & $0.04^{+0.40}_{-0.29}$ & $0.27^{+0.40}_{-0.21}$ & $2318.38^{+2034.19}_{-1119.54}$ & $34.40^{+3.13}_{-5.06}$ \\
\rowcolor{gray!20} IMRPhenomTPHM & $0.21^{+0.21}_{-0.09}$ & $135.87^{+44.76}_{-32.58}$ & $113.00^{+35.71}_{-34.75}$ & $22.90^{+19.84}_{-8.32}$ & $0.78^{+0.16}_{-0.23}$ & $0.55^{+0.39}_{-0.47}$ & $0.56^{+0.18}_{-0.18}$ & $0.36^{+0.31}_{-0.23}$ & $3689.73^{+2897.85}_{-1739.02}$ & $34.35^{+3.44}_{-6.25}$ \\
IMRPhenomXODE & $0.38^{+0.35}_{-0.22}$ & $83.37^{+32.08}_{-16.61}$ & $59.72^{+37.77}_{-18.24}$ & $22.58^{+11.00}_{-8.69}$ & $0.40^{+0.51}_{-0.36}$ & $0.49^{+0.45}_{-0.44}$ & $0.03^{+0.36}_{-0.30}$ & $0.33^{+0.53}_{-0.25}$ & $2210.96^{+1905.15}_{-1082.82}$ & $42.36^{+3.73}_{-5.56}$ \\
\hline\hline
\end{tabular}

\vspace{0.5cm} % spacing between tables

% Second table
\begin{tabular}{lcccccccccc}
\hline\hline
Model & $q$ & $M_{\rm tot}^{\rm source}$ & $m_1^{\rm source}$ & $m_2^{\rm source}$ & $|\chi_1|$ & $|\chi_2|$ & $\chi_{\rm eff}$ & $\chi_{\rm p}$ & $D_{\rm L}$ & $\log\mathcal{L}$\\
\hline
\rowcolor{gray!20} NRSur7dq4 & $0.18^{+0.02}_{-0.01}$ & $236.79^{+17.53}_{-25.14}$ & $201.14^{+15.02}_{-21.30}$ & $35.45^{+3.87}_{-3.90}$ & $0.96^{+0.03}_{-0.07}$ & $0.84^{+0.13}_{-0.34}$ & $-0.60^{+0.22}_{-0.13}$ & $0.60^{+0.21}_{-0.19}$ & $797.93^{+352.14}_{-297.07}$ & $87.81^{+3.97}_{-7.37}$ \\
IMRPhenomXPHM & $0.32^{+0.08}_{-0.10}$ & $254.05^{+25.42}_{-24.79}$ & $192.31^{+27.50}_{-20.76}$ & $61.12^{+12.89}_{-15.65}$ & $0.84^{+0.13}_{-0.19}$ & $0.66^{+0.30}_{-0.53}$ & $-0.34^{+0.14}_{-0.19}$ & $0.61^{+0.15}_{-0.19}$ & $612.57^{+231.56}_{-153.36}$ & $85.81^{+3.97}_{-5.95}$ \\
\rowcolor{gray!20} IMRPhenomTPHM & $0.18^{+0.07}_{-0.05}$ & $216.14^{+35.87}_{-16.37}$ & $183.11^{+32.67}_{-14.81}$ & $32.66^{+13.40}_{-8.26}$ & $0.92^{+0.07}_{-0.22}$ & $0.60^{+0.35}_{-0.51}$ & $-0.77^{+0.26}_{-0.13}$ & $0.24^{+0.28}_{-0.16}$ & $562.57^{+232.14}_{-182.27}$ & $83.12^{+4.38}_{-6.07}$ \\
IMRPhenomXODE & $0.21^{+0.02}_{-0.07}$ & $208.76^{+10.11}_{-14.24}$ & $173.31^{+10.32}_{-9.69}$ & $36.13^{+3.67}_{-12.39}$ & $0.98^{+0.02}_{-0.08}$ & $0.48^{+0.46}_{-0.43}$ & $-0.73^{+0.10}_{-0.09}$ & $0.37^{+0.10}_{-0.05}$ & $365.44^{+238.12}_{-115.05}$ & $100.59^{+5.42}_{-9.83}$ \\
\rowcolor{gray!20} SEOBNRv4PHM & $0.21^{+0.06}_{-0.04}$ & $205.27^{+23.46}_{-81.14}$ & $171.17^{+16.79}_{-69.72}$ & $33.90^{+9.11}_{-12.98}$ & $0.94^{+0.05}_{-0.12}$ & $0.53^{+0.42}_{-0.47}$ & $-0.70^{+0.14}_{-0.11}$ & $0.26^{+0.23}_{-0.18}$ & $582.08^{+229.07}_{-167.74}$ & $-$ \\
\hline\hline
\end{tabular}

\label{tab:combined}
\end{table*}

In this paper, we employ the open-source GW Bayesian inference package \texttt{bilby}~\cite{Ashton:2018jfp} together with the \texttt{NRSur7dq4} waveform model to analyze both events. The strain data for all three detectors are obtained directly from the Gravitational Wave Open Science Center (GWOSC), while the power spectral densities (PSDs) are taken from previous IAS analyses~\cite{Olsen:2022pin,Mehta:2023zlk,Wadekar:2023gea}. For GW190711\_030756, we set the lower and upper frequency cutoffs to $f_{\rm min} = 20\,\mathrm{Hz}$ and $f_{\rm max} = 512\,\mathrm{Hz}$, respectively. For GW200114\_020818, we set $f_{\rm min} = 15\,\mathrm{Hz}$.
For \texttt{NRSur7dq4}, we use a reference frequency of $f_{\rm ref} = 17\,\mathrm{Hz}$. We adopt the same priors as those used in standard LVK analyses and in the GWTC-3 reanalysis employing \texttt{NRSur7dq4} in Ref.~\cite{Islam:2023zzj}. For sampling, we use the nested sampling~\cite{Skilling:2006gxv} algorithm \texttt{dynesty}~\cite{2020MNRAS.493.3132S}. All waveforms are generated using the \texttt{LALSimulation} package~\cite{lalsuite,swiglal}.

To assess waveform systematics, we additionally analyze both events using two other waveform approximants, \texttt{IMRPhenomXPHM} and \texttt{IMRPhenomTPHM}. For \texttt{IMRPhenomTPHM}, we choose a lower reference frequency of $f_{\rm ref} = 10\,\mathrm{Hz}$, as this model prefers a lower reference frequency for waveform generation, whereas for \texttt{IMRPhenomXPHM} we adopt the same reference frequency as in the \texttt{NRSur7dq4} analysis. 
Furthermore, we compare our results with those obtained using \texttt{IMRPhenomXPHM} and \texttt{IMRPhenomXODE} with another open-source software package, \texttt{cogwheel}\footnote{\href{https://github.com/jroulet/cogwheel}{https://github.com/jroulet/cogwheel}}~\cite{Roulet:2024hwz,Islam:2022afg,Roulet:2022kot}, which employs relative binning from Refs.~\cite{Olsen:2022pin,Mehta:2023zlk,Wadekar:2023gea}\footnote{For the \texttt{cogwheel} analyses, we obtain the log-likelihood values from the public data release and apply a PSD drift correction to ensure a consistent comparison between the \texttt{cogwheel} and \texttt{bilby} results.}. Finally, we incorporate publicly available posterior samples from Ref.~\cite{Ruiz-Rocha:2025yno} for \texttt{SEOBNRv4PHM} for GW200114\_020818 obtained using a third parameter-estimation framework, \texttt{RIFT}~\cite{Pankow:2015cra,Lange:2018pyp}. We do not perform independent parameter-estimation analyses with \texttt{SEOBNRv4PHM} or its successors, as these waveform models are computationally expensive.

In the following sections, we summarize our parameter-estimation results using \texttt{NRSur7dq4} (source properties in Section~\ref{sec:results} and remnant properties in Section~\ref{sec:remnants}) with an additional focus on understanding waveform systematics. We then discuss their astrophysical implications in the context of other GW events detected to date, as well as in light of N-body (and few-body) star-cluster simulations in Section~\ref{sec:implications}. We have made our results publicly available at \href{https://github.com/tousifislam/GW190711\_GW200114}{https://github.com/tousifislam/GW190711\_GW200114}.

%%%%%%%%%%%%%%%%%%%%%%%%%%%%%%%%%%%%%%%%%%%%%%%%%%%%%%%%%%%%%%%%%%%%%%%%%%%%%%%%%%%
%%%%%%%%%%%%%%%%%%%%%%%%%%%%%%%%%%%%%%%%%%%%%%%%%%%%%%%%%%%%%%%%%%%%%%%%%%%%%%%%%%%
\section{Estimation of the source properties}
\label{sec:results}
%%%%%%%%%%%%%%%%%%%%%%%%%%%%%%%%%%%%%%%%%%%%%%%%%%%%%%%%%%%%%%%%%%%%%%%%%%%%%%%%%%%
%%%%%%%%%%%%%%%%%%%%%%%%%%%%%%%%%%%%%%%%%%%%%%%%%%%%%%%%%%%%%%%%%%%%%%%%%%%%%%%%%%%
For each event, we infer all 15 BBH source parameters that describe a quasi-circular BBH merger. These consist of eight intrinsic and seven extrinsic parameters. The intrinsic parameters include the component masses $m_1$ and $m_2$, the dimensionless spin magnitudes $|\chi_1|$ and $|\chi_2|$, the spin tilt angles $\theta_1$ and $\theta_2$ measured relative to the orbital angular momentum, and the azimuthal spin angles $\phi_{12}$ and $\phi_{JL}$, where $L$ denotes the orbital angular momentum and $J$ denotes the total angular momentum. The extrinsic parameters describe the source location and orientation relative to the detector. The angle between the total angular momentum of the binary and the line of sight to the detector is denoted by $\iota$, often referred to as the inclination angle, while the luminosity distance to the source is denoted by $D_L$. The right ascension $\alpha$ and declination $\delta$ specify the sky location of the source, $\psi$ denotes the polarization angle, $\phi_c$ is the coalescence phase, and $t_c$ denotes the coalescence time. For details of these parameters, we refer the readers to Refs.~\cite{Ashton:2018jfp,Veitch:2014wba}.
Additionally, we compute two effective spin quantities: the effective inspiral spin parameter~\cite{Ajith:2009bn,Santamaria:2010yb,Vitale:2016avz}
\begin{equation}
\chi_{\rm eff} = \frac{m_1 \chi_1 \cos\theta_1 + m_2 \chi_2 \cos\theta_2}{m_1 + m_2},
\label{eq:chieff}
\end{equation}
and the effective transverse spin precession parameter~\cite{Hannam:2013oca,Schmidt:2014iyl}
\begin{equation}
\chi_{\rm p} = \max \left( \chi_1 \sin\theta_1,\; \frac{q(4q+3)}{4+3q}\,\chi_2 \sin\theta_2 \right).
\label{eq:chip}
\end{equation}
Finally, we denote the total mass as $M_{\rm tot}=m_1+m_2$. When discussing the masses, we use the superscript ``det'' to denote the detector frame and ``source'' to denote the source frame. The two are related as $M_{\rm tot}^{\rm det} = (1+z)\,M_{\rm tot}^{\rm source}$ where $z$ is the cosmological redshift.

%%%%%%%%%%%%%%%%%%%%%%%%%%%%%%%%%%%%%%%%%%%%%%%%%%%%%%%%%%%%%%%%%%%%%%%%%%%%%%%%%%%
\begin{figure*}
    \centering
    \includegraphics[width=0.95\textwidth]{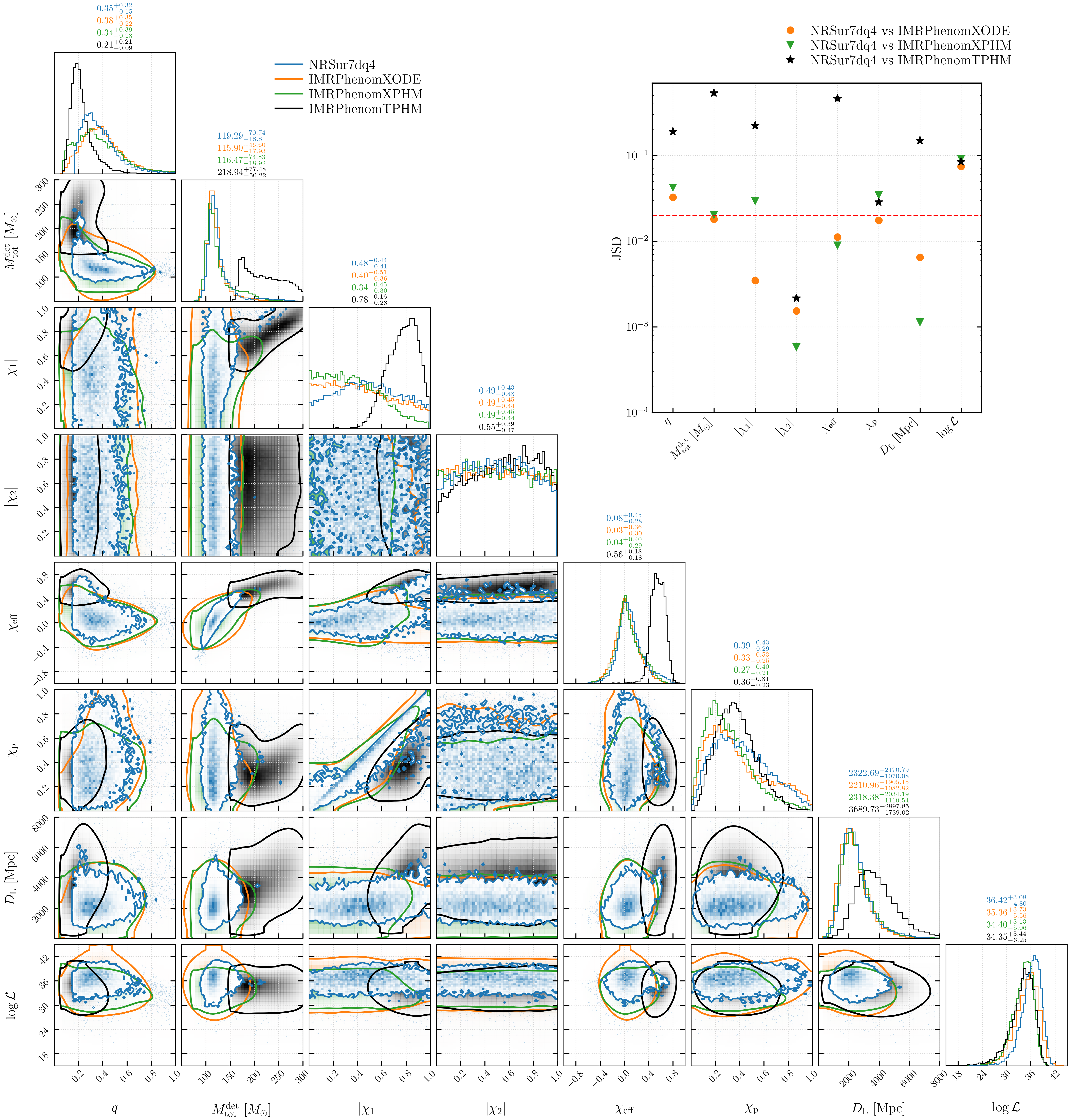}
    \caption{We show the posteriors for eight key source properties of GW190711\_030756 ($p_\mathrm{astro}=0.99$) inferred using four different waveform models: \texttt{NRSur7dq4} (blue), \texttt{IMRPhenomXODE} (orange), \texttt{IMRPhenomXPHM} (green), and \texttt{IMRPhenomTPHM} (black). In addition, we report the Jensen--Shannon divergence between \texttt{NRSur7dq4} and each of the other waveform models for all parameters (see inset). Note that we restrict the \texttt{NRSur7dq4} approximant to $q \geq 0.167$ to remain within its validation range throughout this paper. More details are in Section~\ref{sec:wavesys}.}
    \label{fig:gw190711_posteriors}
\end{figure*}
%%%%%%%%%%%%%%%%%%%%%%%%%%%%%%%%%%%%%%%%%%%%%%%%%%%%%%%%%%%%%%%%%%%%%%%%%%%%%%%%%%%

%%%%%%%%%%%%%%%%%%%%%%%%%%%%%%%%%%%%%%%%%%%%%%%%%%%%%%%%%%%%%%%%%%%%%%%%%%%%%%%%%%%
\begin{figure*}
    \centering
    \includegraphics[width=0.95\textwidth]{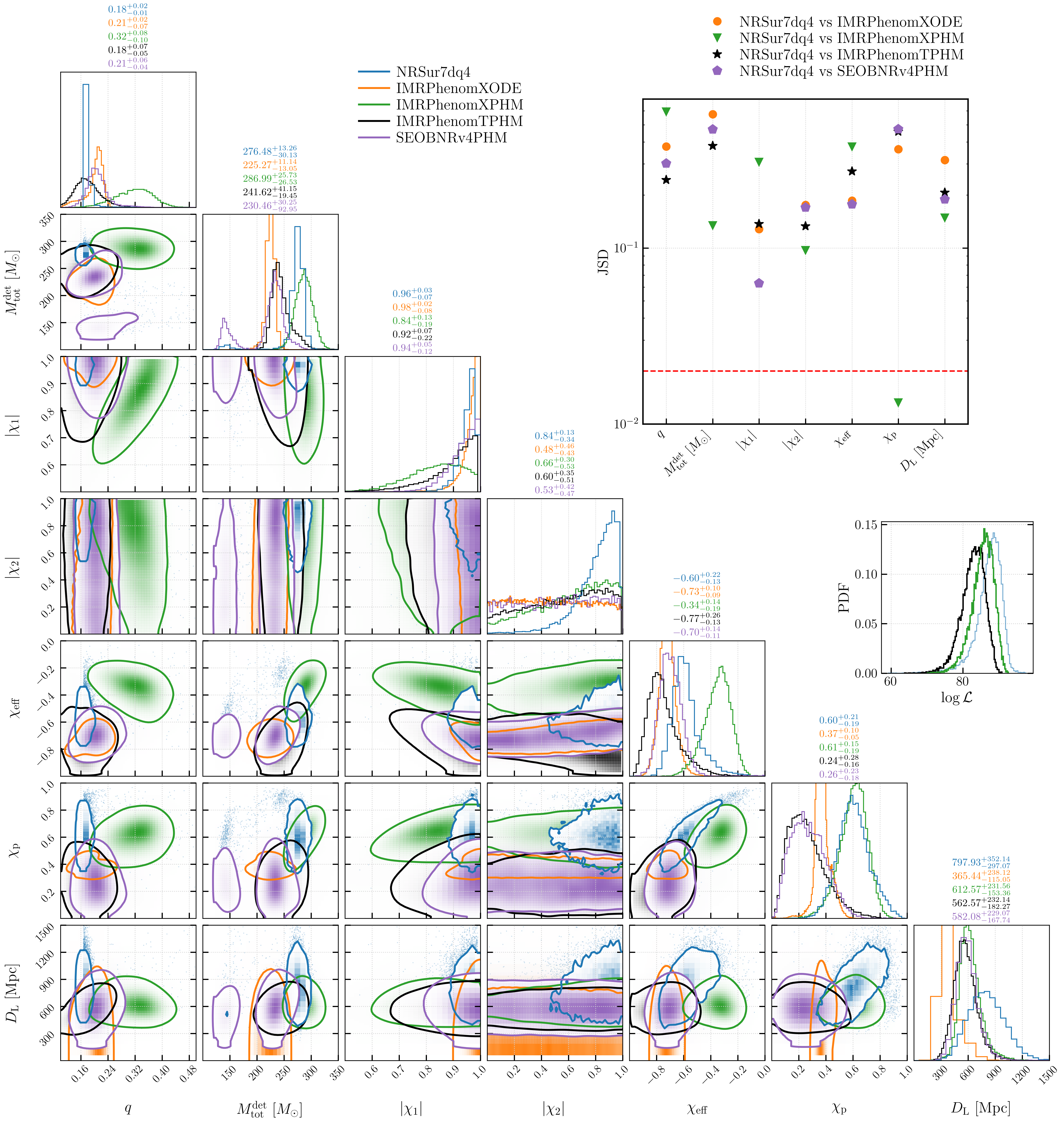}
    \caption{Similar to Fig.~\ref{fig:gw190711_posteriors} but instead for the properties of GW200114\_020818. Unlike GW190711\_030756, this event has marginal significance with $p_\mathrm{astro}=0.71$. Note, however, that the $p_\mathrm{astro}$ value is sensitive to the assumed astrophysical population prior in the search, which remains poorly constrained in the high-mass regime.
    More details are in Section~\ref{sec:wavesys}.}
    \label{fig:gw200114_posteriors}
\end{figure*}
%%%%%%%%%%%%%%%%%%%%%%%%%%%%%%%%%%%%%%%%%%%%%%%%%%%%%%%%%%%%%%%%%%%%%%%%%%%%%%%%%%%

%%%%%%%%%%%%%%%%%%%%%%%%%%%%%%%%%%%%%%%%%%%%%%%%%%%%%%%%%%%%%%%%%%%%%%%%%%%%%%%%%%%
\subsection{Results from \texttt{NRSur7dq4}}
\label{sec:nrsurpe}
%%%%%%%%%%%%%%%%%%%%%%%%%%%%%%%%%%%%%%%%%%%%%%%%%%%%%%%%%%%%%%%%%%%%%%%%%%%%%%%%%%%
First we focus on \texttt{NRSur7dq4} posteriors. 
We present the \texttt{NRSur7dq4} results for GW190711\_030756 (dark blue posteriors) and GW200114\_020818 (crimson posteriors) together in Fig.~\ref{fig:combined_posteriors}. Posteriors for more parameters are shown in Figs.~\ref{fig:gw190711_posteriors} and~\ref{fig:gw200114_posteriors}.
We show the inferred posterior distributions for five key intrinsic source parameters: the inverse mass ratio $1/q$, the dimensionless spin magnitudes $|\chi_{1,2}|$, the effective inspiral spin parameter $\chi_{\rm eff}$, and the effective transverse spin precession parameter $\chi_{\rm p}$. 
A summary of the inferred source parameters (which includes the source-frame total mass $M_{\rm tot}^{\rm source}$ (in $M_{\odot}$), individual source-frame masses $m_{1,2}^{\rm source}$ (in $M_{\odot}$), and the luminosity distance $D_L$ (in Mpc)) is also provided in Table~\ref{tab:combined}.
For reference, we also show the posteriors for GW231123\_135430, one of the most massive and spinning GW events so far, together with the available SXS NR simulations within $q \in [0.1667,1]$ (along with simulations from the RIT and MAYA catalogs). A subset of these NR simulations has been used to train the \texttt{NRSur7dq4} model. 
An even smaller subset has been used to test the semi-analytical precessing-spin 
waveform models employed in this work. We find that the inferred spin-magnitude posteriors of GW200114\_020818 and GW231123\_135430 have little overlap with the existing NR data coverage. In other words, the inferred parameters of both events lie in an extreme corner of the currently explored parameter space. On the other hand, the GW190711\_030756 posteriors lie within the region of parameter space covered by the NR simulations. These events therefore provide a testbed for exploring possible waveform systematics.

In Fig.~\ref{fig:whitened_strain}, we show the whitened strain data for GW190711\_030756 and GW200114\_020818 as observed in the H1 and L1 detectors, together with the best-fit template obtained using the \texttt{NRSur7dq4} model. In addition, we overlay 1000 waveform templates corresponding to randomly selected source parameters drawn from the inferred \texttt{NRSur7dq4} posterior.

We find that the data favor an interpretation of both GW190711\_030756 and GW200114\_020818 as unequal-mass binaries as seen in Figs.~\ref{fig:gw190711_posteriors} and~\ref{fig:gw200114_posteriors}.
However, this interpretation requires some care. The \texttt{NRSur7dq4} model is trained within the range $0.25 \leq q \leq 1$ and validated down to $0.167 \leq q \leq 1$, and we therefore impose a hard prior boundary at $q = 0.167$. While the mass-ratio posterior for GW190711\_030756 is well contained within this range, GW200114\_020818 exhibits strong prior railing at the lower boundary $q = 0.167$. This behavior indicates either that \texttt{NRSur7dq4} finds no viable solution for this event within the range $0.167 \leq q \leq 1$, or that the preferred posterior support for $q$ peaks at the prior boundary. Consequently, the inferred value of $q$ for GW200114\_020818, as well as other correlated parameters, should be interpreted conservatively within the domain of model validity. However, as we examine waveform systematics using additional waveform models in later sections, we find consistent evidence that the true solution likely lies close to the boundary at $q = 0.167$. In either case, a robust inference is that both events strongly favor unequal-mass configurations.

We infer source-frame total masses of $85.43^{+35.71}_{-17.37}\,M_{\odot}$ for GW190711\_030756 and $234.21^{+18.33}_{-19.98}\,M_{\odot}$ for GW200114\_020818, indicating that both systems are unusually massive. Consequently, the primary black hole in GW190711\_030756 lies within the pair-instability supernova (PISN) mass gap~\cite{Woosley:2021xba}, while the primary component of GW200114\_020818 is inferred to be an intermediate-mass black hole (IMBH) i.e. black holes with masses or $100M_{\odot}$ or beyond. Furthermore, the source-frame total mass of GW200114\_020818 is quite similar to GW231123\_135430 ($\sim 230M_{\odot}$).

Next, we focus on the spin parameters. GW190711\_030756 does not exhibit any particularly distinctive spin features, with the inferred effective inspiral spin $\chi_{\rm eff}$ consistent with zero and the effective precession parameter $\chi_{\rm p}$ largely unconstrained, that is, consistent with the prior. The dimensionless spin magnitudes of both black holes are also weakly constrained. 

In contrast, GW200114\_020818 exhibits striking features across nearly all spin parameters. We find that both black holes are inferred to be rapidly spinning, with dimensionless spin magnitudes of $|\chi_1| = 0.96^{+0.03}_{-0.07}$ and $|\chi_2| = 0.87^{+0.11}_{-0.30}$. Furthermore, the spins are preferentially anti-aligned with the orbital angular momentum, resulting in a strongly negative effective inspiral spin of $\chi_{\rm eff} = -0.56^{+0.24}_{-0.15}$. In addition, the system exhibits significant spin-induced precession, characterized by an effective precession parameter of $\chi_{\rm p} = 0.65^{+0.21}_{-0.21}$.

Finally, we examine the inferred distances. We find luminosity distances of $2322^{+2170}_{-1070}\,\mathrm{Mpc}$ and $854^{+350}_{-328}\,\mathrm{Mpc}$ for GW190711\_030756 and GW200114\_020818, respectively. These correspond to redshifts of $0.41^{+0.30}_{-0.17}$ and $0.17^{+0.06}_{-0.06}$ (using Planck 2018 $\Lambda$CDM cosmology~\cite{Planck2018VI}). 

We note that GW200114\_020818 has previously been analyzed using the \texttt{NRSur7dq4} waveform model in conjunction with the \texttt{RIFT} inference framework~\cite{Ruiz-Rocha:2025yno}. Our results show significant differences with that earlier analysis. We attribute these discrepancies primarily to differences in the assumed priors and sampling methodologies. In particular, Ref.~\cite{Ruiz-Rocha:2025yno} imposed a prior cut of $|\chi_{1,2}| \leq 0.8$, which restricts the allowed spin magnitudes, and employed a different sampling strategy.

Finally, using the dynamics surrogate of \texttt{NRSur7dq4}, we investigate the binary dynamics of these two events and examine three possible signatures of nontrivial spin behavior: (i) up--down instability~\cite{Gerosa:2015hba,Lousto:2016nlp,Johnson-McDaniel:2021rvv,Varma:2020bon}, (ii) transitional precession~\cite{PhysRevD.49.6274}, and (iii) spin--orbit resonance~\cite{Schnittman:2004vq}. We do not find evidence for any of these scenarios. Details of this investigation are presented in Appendix~\ref{sec:dynamics}.

%%%%%%%%%%%%%%%%%%%%%%%%%%%%%%%%%%%%%%%%%%%%%%%%%%%%%%%%%%%%%%%%%%%%%%%%%%%%%%%%%%%
\subsection{Waveform systematics}
\label{sec:wavesys}
%%%%%%%%%%%%%%%%%%%%%%%%%%%%%%%%%%%%%%%%%%%%%%%%%%%%%%%%%%%%%%%%%%%%%%%%%%%%%%%%%%%
Next, we investigate waveform systematics for GW190711\_030756 and GW200114\_020818. For both events, we compare results obtained with \texttt{NRSur7dq4}, \texttt{IMRPhenomXPHM}, and \texttt{IMRPhenomTPHM} using \texttt{bilby}. In addition, we include results obtained with \texttt{IMRPhenomXODE} using \texttt{cogwheel}. For GW200114\_020818, we further incorporate \texttt{SEOBNRv4PHM} results obtained with \texttt{RIFT}.
To quantify waveform systematics, we visually compare the posteriors from different waveform models and check the recovered maximum log-likelihood values from these analyses. For key source properties, we also compute the Jensen--Shannon divergence (JSD)~\cite{Lin1991} between the marginalized one-dimensional posteriors obtained with \texttt{NRSur7dq4} and those from the other waveform approximants. The JSD is symmetric and bounded, $0 \le \mathrm{JSD}(p \,\|\, q) \le \log 2 \approx 0.693$. Values of $\mathrm{JSD} \lesssim 0.02$ indicate statistically consistent distributions, while $\mathrm{JSD} \gtrsim 0.1$ signify substantial differences; intermediate values correspond to moderately distinct posterior distributions.

In Figs.~\ref{fig:gw190711_posteriors} and~\ref{fig:gw200114_posteriors}, we show the posterior distributions obtained with different waveform models for GW190711\_030756 and GW200114\_020818, respectively. A summary of the inferred parameters for all waveform models are also provided in Table~\ref{tab:combined}. The first thing we confirm is that, for both events, \texttt{NRSur7dq4} yields the highest recovered likelihood, indicating that its inferred parameters are the most reliable among the models considered.

\begin{figure}
    \centering
    \includegraphics[width=\columnwidth]{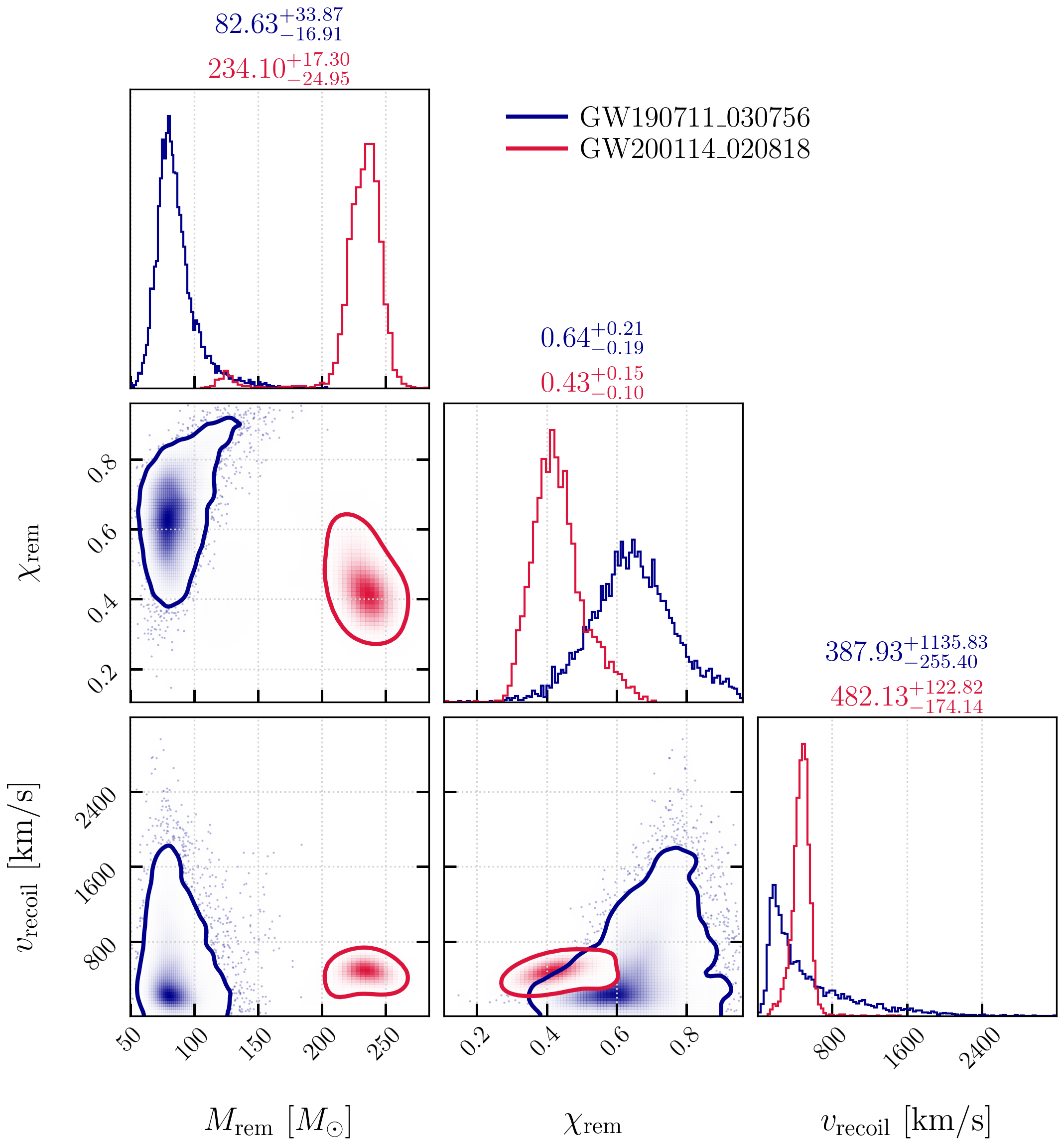}
    \caption{We show the posteriors for the remnant mass, spin and recoil kick velocity for GW190711\_030756 (dark blue) and GW200114\_020818 (crimson), inferred using the numerical-relativity precessing surrogate remnant model \texttt{NRSur7dq4Remnant}. More details are in Section~\ref{sec:remnants}.}
    \label{fig:combined_remnant_posteriors}
\end{figure}

For GW190711\_030756, we find broad agreement among most waveform models, while \texttt{IMRPhenomTPHM} exhibits significant deviations from the other models across nearly all parameters (Fig.~\ref{fig:gw190711_posteriors}). In particular, the JSD values between \texttt{IMRPhenomXPHM} or \texttt{IMRPhenomXODE} and \texttt{NRSur7dq4} are close to or below the threshold of $0.02$ for most parameters, with the exception of $|\chi_2|$. In contrast, the JSD values between \texttt{IMRPhenomTPHM} and \texttt{NRSur7dq4} exceed this threshold for several parameters and can be as large as $0.5$, indicating substantial discrepancies between these models. We note that the \texttt{IMRPhenomTPHM} model favors a configuration with a positive effective inspiral spin, rather than one consistent with zero. It also prefers a more massive binary compared to the other waveform models and infers a significantly spinning primary black hole.

For GW200114\_020818, waveform systematics are substantial, with almost no two models yielding consistent results (Fig.~\ref{fig:gw200114_posteriors}). For nearly all parameters, the JSD values between the \texttt{NRSur7dq4} posteriors and those obtained with other waveform models is significantly above the threshold for statistical consistency and is close to $0.1$. In this sense, GW200114\_020818 exhibits behavior reminiscent of GW231123\_135430. Several notable differences emerge across the models. First, most waveform models favor an asymmetric binary configuration with mass ratios $q \lesssim 0.25$, whereas \texttt{IMRPhenomXPHM} prefers mass ratios closer to the equal-mass regime near $q\sim 0.3$. Second, the secondary spin magnitude $|\chi_2|$ remains largely unconstrained for all models except \texttt{NRSur7dq4}, which infers a significantly spinning secondary black hole. Third, although the models disagree on the precise value of the effective inspiral spin, all favor confidently negative values. Finally, while \texttt{NRSur7dq4} and \texttt{IMRPhenomXPHM} infer significant spin-induced precession, \texttt{IMRPhenomTPHM} and \texttt{SEOBNRv4PHM} favor configurations with only modest precession. 
A potential explanation for the strong waveform systematics associated with GW200114\_020818 is the growing divergence among waveform models in the high--mass-ratio, high-spin regime. Semi-analytical models, in particular, may suffer substantial loss of accuracy as the system moves away from the near-equal-mass limit~\cite{Pratten:2020ceb,Yu:2023lml,MacUilliam:2024oif}. Additionally, the extremal spin magnitudes inferred for GW200114\_020818 fall outside the training domain of \texttt{NRSur7dq4}.

\begin{figure}
    \centering
    \includegraphics[width=\columnwidth]{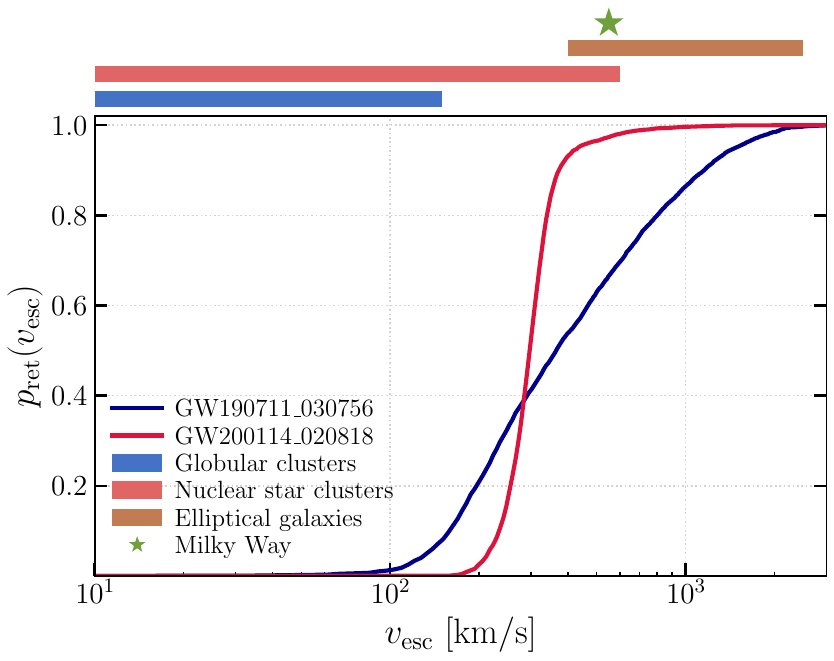}
    \caption{Retention probability of remnant black hole for GW190711\_030756 (dark blue) and GW200114\_020818 (crimson) as a function of the host environment's escape velocity. Typical escape-velocity ranges for globular clusters (blue bar), nuclear star clusters (red bar), elliptical galaxies (brown bar), and the Milky Way (green star) are indicated for reference. More details are in Section~\ref{sec:retention}.}
    \label{fig:recoil_implications}
\end{figure}

%%%%%%%%%%%%%%%%%%%%%%%%%%%%%%%%%%%%%%%%%%%%%%%%%%%%%%%%%%%%%%%%%%%%%%%%%%%%%%%%%%%
%%%%%%%%%%%%%%%%%%%%%%%%%%%%%%%%%%%%%%%%%%%%%%%%%%%%%%%%%%%%%%%%%%%%%%%%%%%%%%%%%%%
\section{Estimation of the remnant properties}
\label{sec:remnants}
%%%%%%%%%%%%%%%%%%%%%%%%%%%%%%%%%%%%%%%%%%%%%%%%%%%%%%%%%%%%%%%%%%%%%%%%%%%%%%%%%%%
%%%%%%%%%%%%%%%%%%%%%%%%%%%%%%%%%%%%%%%%%%%%%%%%%%%%%%%%%%%%%%%%%%%%%%%%%%%%%%%%%%%
We now take the inferred source parameters obtained using the \texttt{NRSur7dq4} waveform model and propagate them through \texttt{NRSur7dq4Remnant}~\cite{Varma:2019csw} to compute the corresponding remnant properties following Ref.~\cite{Islam:2023zzj}. We access \texttt{NRSur7dq4Remnant} model through \texttt{surfinBH} package~\footnote{\href{https://github.com/vijayvarma392/surfinBH}{https://github.com/vijayvarma392/surfinBH}}.
We again restrict our analysis to $q\geq 0.167$ to be within the validation domain of both \texttt{NRSur7dq4Remnant} and \texttt{NRSur7dq4}.
We focus on the remnant mass, spin, and recoil kick velocity magnitude. Schematically, we can write the process as:
\begin{equation}
    \{m_1, m_2, |\chi_1|, |\chi_2|, \theta_1, \theta_2, \phi_1, \phi_2\} \mapsto \{M_{\rm rem},\chi_{\rm rem},v_{\rm recoil}\}.
\end{equation}
We show the inferred renmant properties in Fig.~\ref{fig:combined_remnant_posteriors}.
We infer the remnant black hole mass for GW190711\_030756 to be $82.6^{+33.8}_{-16.9}\,M_{\odot}$, placing it within the pair-instability supernova mass gap. For GW200114\_020818, the remnant black hole mass is inferred to be $231.2^{+17.9}_{-19.4}\,M_{\odot}$, firmly in the IMBH regime.
The inferred remnant spin magnitudes for GW190711\_030756 and GW200114\_020818 are $0.64^{+0.19}_{-0.21}$ and $0.46^{+0.11}_{-0.15}$, respectively. The combination of significant spin precession and a more asymmetric mass ratio in GW200114\_020818 leads to a slightly smaller remnant spin compared to GW190711\_030756. Finally, we compute the recoil kick velocities. For GW190711\_030756, we obtain a broad posterior that closely follows the prior, with an inferred kick velocity of $365^{+1100}_{-232}\,\mathrm{km\,s^{-1}}$.
We note that, while the use of semi-analytical models for the remnant mass and spin would have a negligible impact on our results, the inferred recoil kick velocities could change appreciably. However, \texttt{NRSur7dq4Remnant} has been shown to provide higher accuracy than existing semi-analytical kick models~\cite{Lousto:2008dn,Lousto:2010xk,Lousto:2012gt,Lousto:2012su}, and we therefore do not consider alternative prescriptions.

\begin{figure}
    \centering
    \includegraphics[width=0.4\textwidth]{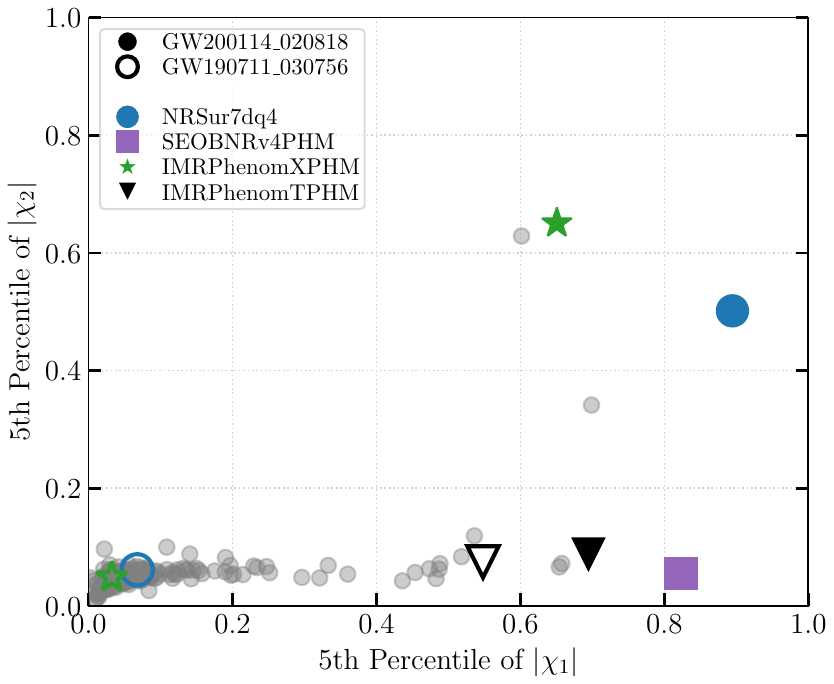}
    \includegraphics[width=0.4\textwidth]{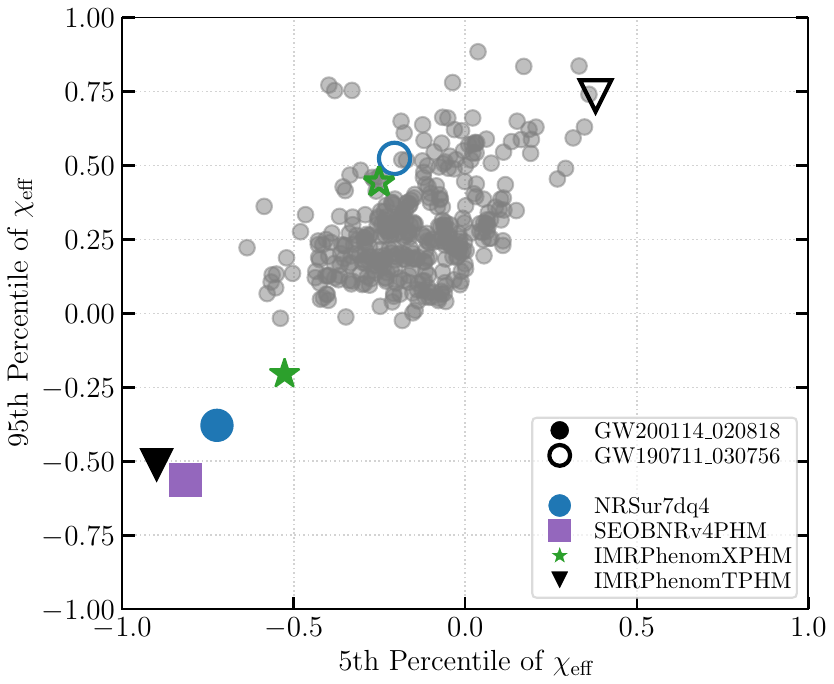}
    \includegraphics[width=0.4\textwidth]{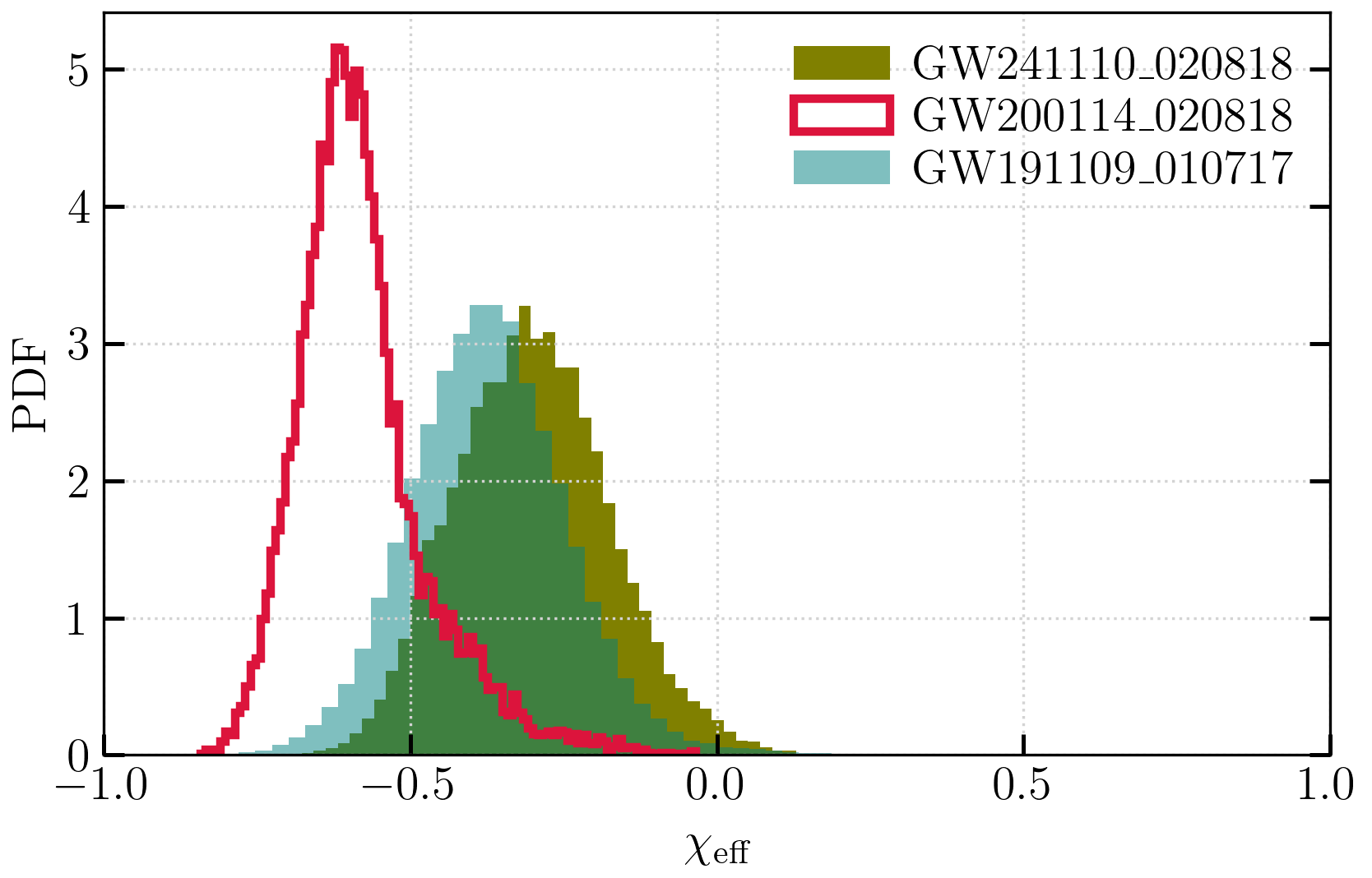}
    \caption{\textbf{Top:} We show the 5th percentiles of the inferred dimensionless spin magnitudes $|\chi_1|$ and $|\chi_2|$ for all GWTC-4 events obtained using different waveform approximants, shown as gray circles. For comparison, we overlay the values inferred from four different waveform models for GW190711\_030756 (open markers) and GW200114\_020818 (filled markers). We find that the 5th-percentile spin magnitudes inferred with \texttt{NRSur7dq4} for GW200114\_020818 are unusually high and represent clear outliers relative to the rest of the catalog.
    \textbf{Middle:} Same as the top panel, but showing the 5th and 95th percentiles of $\chi_{\rm eff}$. As before, the values inferred using \texttt{NRSur7dq4} for GW200114\_020818 are extreme on the negative side compared to all other events. More details are in Section~\ref{sec:formation_gw19} and Section~\ref{sec:exception_gw20}. \textbf{Bottom}: We show the comparison of $\chi_{\rm eff}$ for GW200114\_020818 with the two events from GWTC-4 having the most negative $\chi_{\rm eff}$ values (GW241110\_043853 and GW191109\_010717). More details are in Section~\ref{sec:exception_gw20}.}
    \label{fig:spin_percentiles}
\end{figure}

% \begin{figure}
%     \centering
%     \includegraphics[width=\columnwidth]{gw200114_chieff_comparison.png}
%     \caption{We show the inferred negative effective inspiral spin $\chi_{\rm eff}$ for GW200114\_020818. For comparison, we also include the two events (GW241110\_043853 and GW191109\_010717) from the GWTC-4 catalog with the next most negative inferred values of $\chi_{\rm eff}$. More details are in Section~\ref{sec:exception_gw20}.}
%     \label{fig:gw200114_chieff_comparison}
% \end{figure}

%%%%%%%%%%%%%%%%%%%%%%%%%%%%%%%%%%%%%%%%%%%%%%%%%%%%%%%%%%%%%%%%%%%%%%%%%%%%%%%%%%%
%%%%%%%%%%%%%%%%%%%%%%%%%%%%%%%%%%%%%%%%%%%%%%%%%%%%%%%%%%%%%%%%%%%%%%%%%%%%%%%%%%%
\section{Astrophysical implications}
\label{sec:implications}
%%%%%%%%%%%%%%%%%%%%%%%%%%%%%%%%%%%%%%%%%%%%%%%%%%%%%%%%%%%%%%%%%%%%%%%%%%%%%%%%%%%
%%%%%%%%%%%%%%%%%%%%%%%%%%%%%%%%%%%%%%%%%%%%%%%%%%%%%%%%%%%%%%%%%%%%%%%%%%%%%%%%%%%
Finally, we discuss the astrophysical implications of our results in Secs.~\ref{sec:results} and~\ref{sec:remnants}. As \texttt{NRSur7dq4} yields the highest likelihood for both events, we treat \texttt{NRSur7dq4} as the reference model for astrophysical interpretation unless otherwise specified.

%%%%%%%%%%%%%%%%%%%%%%%%%%%%%%%%%%%%%%%%%%%%%%%%%%%%%%%%%%%%%%%%%%%%%%%%%%%%%%%%%%%
\subsection{Retention of the remnant black hole}
\label{sec:retention}
%%%%%%%%%%%%%%%%%%%%%%%%%%%%%%%%%%%%%%%%%%%%%%%%%%%%%%%%%%%%%%%%%%%%%%%%%%%%%%%%%%%
We begin by focusing on the remnant black holes and assess the probability that they are retained in their host environments. To estimate this, we adopt representative astrophysical ranges for the escape velocities of different environments: globular clusters ($0$--$150\,\mathrm{km\,s^{-1}}$), nuclear star clusters ($0$--$600\,\mathrm{km\,s^{-1}}$), and massive elliptical galaxies ($400$--$2500\,\mathrm{km\,s^{-1}}$)~\cite{Merritt:2004xa,Antonini:2016gqe}. In Fig.~\ref{fig:recoil_implications}, we show the retention probability of the remnant black holes for both events as a function of the escape velocity.
We then find that the \textit{overall} probability for the merger remnant to be retained in its host environment, and thus potentially participate in subsequent mergers with other black holes, is $0.079$, $0.62$, and $0.997$ for GW190711\_030756 if the merger occurred in a globular cluster, a nuclear star cluster, or an elliptical galaxy, respectively. For GW200114\_020818, the corresponding retention probabilities are $0.0002$, $0.965$, and $1$ in globular clusters, nuclear star clusters, and elliptical galaxies.

%%%%%%%%%%%%%%%%%%%%%%%%%%%%%%%%%%%%%%%%%%%%%%%%%%%%%%%%%%%%%%%%%%%%%%%%%%%%%%%%%%%
\subsection{Formation channel for GW190711\_030756}
\label{sec:formation_gw19}
%%%%%%%%%%%%%%%%%%%%%%%%%%%%%%%%%%%%%%%%%%%%%%%%%%%%%%%%%%%%%%%%%%%%%%%%%%%%%%%%%%%
The spin measurements for GW190711\_030756 are largely unremarkable (Fig.~\ref{fig:spin_percentiles}). 
Here, in the upper panel, we plot the 5th percentiles of the inferred dimensionless spin magnitudes $|\chi_1|$ and $|\chi_2|$ for all GWTC events obtained using different waveform approximants, shown as gray circles. For comparison, we overlay the values inferred from four different waveform models for GW190711\_030756.
With the exception of \texttt{IMRPhenomTPHM}, all waveform models considered, including \texttt{NRSur7dq4}, yield weakly constrained spin magnitudes $|\chi_{1,2}|$. The effective inspiral spin is consistent with zero and broadly consistent with the prior, while the effective precession parameter remains small and similarly prior dominated (Fig.~\ref{fig:combined_posteriors}). 
Two parameters, however, warrant closer attention. The inferred mass ratio $q = 0.35^{+0.32}_{-0.15}$, indicates a moderately asymmetric system, and the primary BH mass lies within the PISN mass gap. Together, these features suggest a likely dynamical origin of the binary, potentially in a dense stellar cluster environment.

\begin{figure*}
    \centering
    \includegraphics[width=0.48\textwidth]{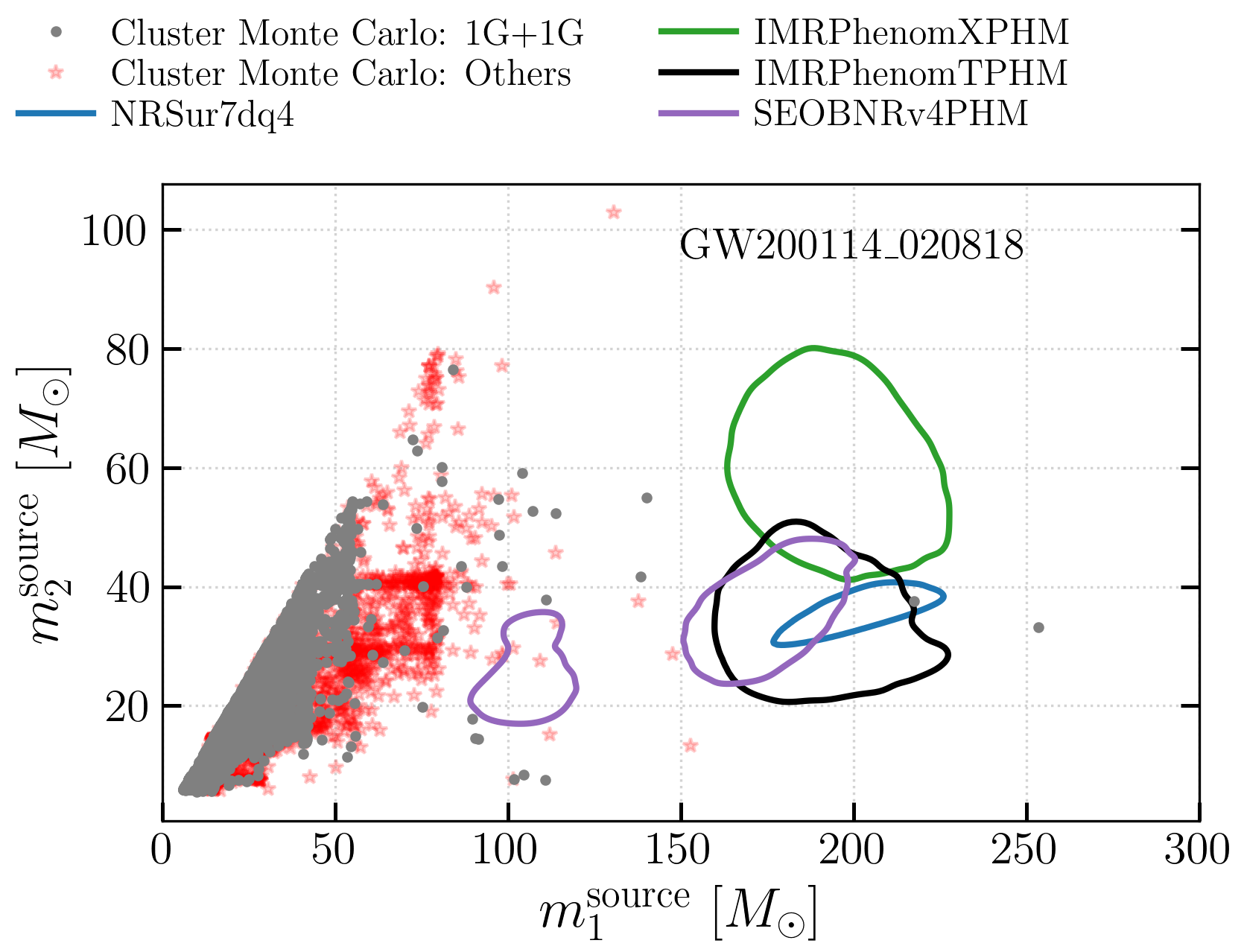}
    \includegraphics[width=0.48\textwidth]{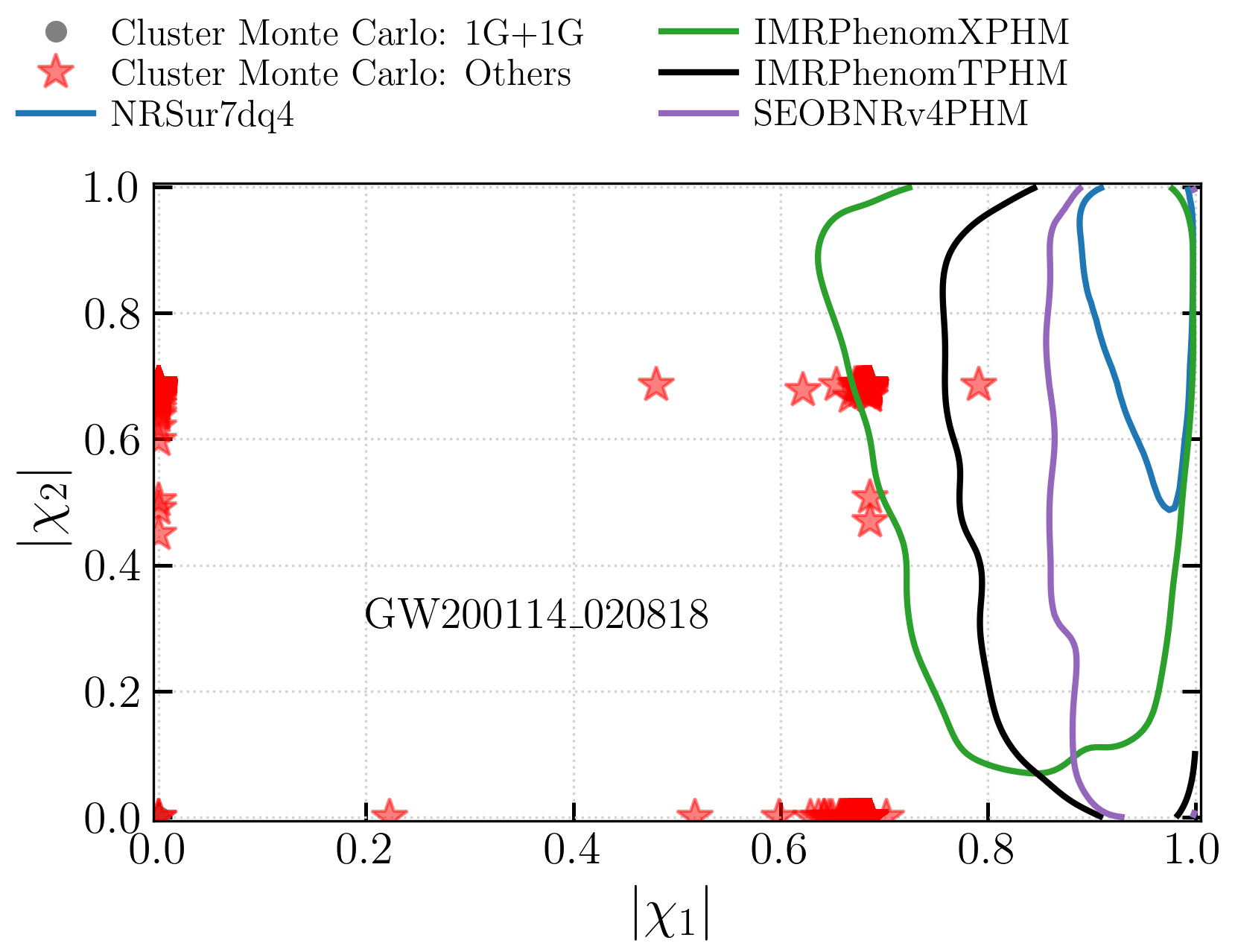}
    \caption{
    \textbf{Left panel:} We show the source-frame component masses $m_1^{\rm source}$ and $m_2^{\rm source}$ of all BBH mergers formed dynamically in globular clusters as found in the \texttt{CMC} catalog. For comparison, we overlay the inferred 90\% contour of source-frame masses of GW200114\_020818 obtained using different waveform models.
    \textbf{Right panel:} Same as the left panel, but showing the dimensionless spin magnitudes $|\chi_1|$ and $|\chi_2|$. More details are in Section~\ref{sec:formation_gw20}.
    }
    \label{fig:cmc}
\end{figure*}

%%%%%%%%%%%%%%%%%%%%%%%%%%%%%%%%%%%%%%%%%%%%%%%%%%%%%%%%%%%%%%%%%%%%%%%%%%%%%%%%%%%
\subsection{Exceptional spin inference for GW200114\_020818}
\label{sec:exception_gw20}
%%%%%%%%%%%%%%%%%%%%%%%%%%%%%%%%%%%%%%%%%%%%%%%%%%%%%%%%%%%%%%%%%%%%%%%%%%%%%%%%%%%
Next, we focus on the exceptional nature of the spin measurements for GW200114\_020818. 
In particular, we investigate how distinct the inferred spin magnitudes $|\chi_1|$ and $|\chi_2|$, as well as the effective inspiral spin $\chi_{\rm eff}$, are for this event in comparison with all other events detected to date in the GWTC catalog. 
In Fig.~\ref{fig:spin_percentiles} (\textit{upper panel}), we plot the 5th percentiles of the inferred dimensionless spin magnitudes $|\chi_1|$ and $|\chi_2|$ for all GWTC events and overlay the values inferred from four different waveform models for GW200114\_020818 (filled markers). We find that the 5th-percentile spin magnitudes inferred with \texttt{NRSur7dq4} for GW200114\_020818 are unusually high and constitute clear outliers relative to the rest of the catalog.

For the effective inspiral spin $\chi_{\rm eff}$, we show both the 5th and 95th percentiles (Fig.~\ref{fig:spin_percentiles}; \textit{middle panel}). Consistent with the behavior seen in the spin magnitudes, the values inferred using \texttt{NRSur7dq4} for GW200114\_020818 is a strong outlier compared with all other GWTC events.
In particular, GW200114\_020818 exhibits the most negative effective inspiral spin $\chi_{\rm eff}$ among all GW events and candidates identified to date. In the bottom panel, we show the $\chi_{\rm eff}$ posterior for GW200114\_020818 together with the posteriors for the two events from the GWTC-4 catalog with the next most negative inferred values, GW241110\_043853 and GW191109\_010717. While the posteriors for GW241110\_043853 and GW191109\_010717 have some support at positive values of $\chi_{\rm eff}$, the posterior for GW200114\_020818 lies almost entirely in the negative $\chi_{\rm eff}$ region.

\begin{figure}
    \centering
    \includegraphics[width=\columnwidth]{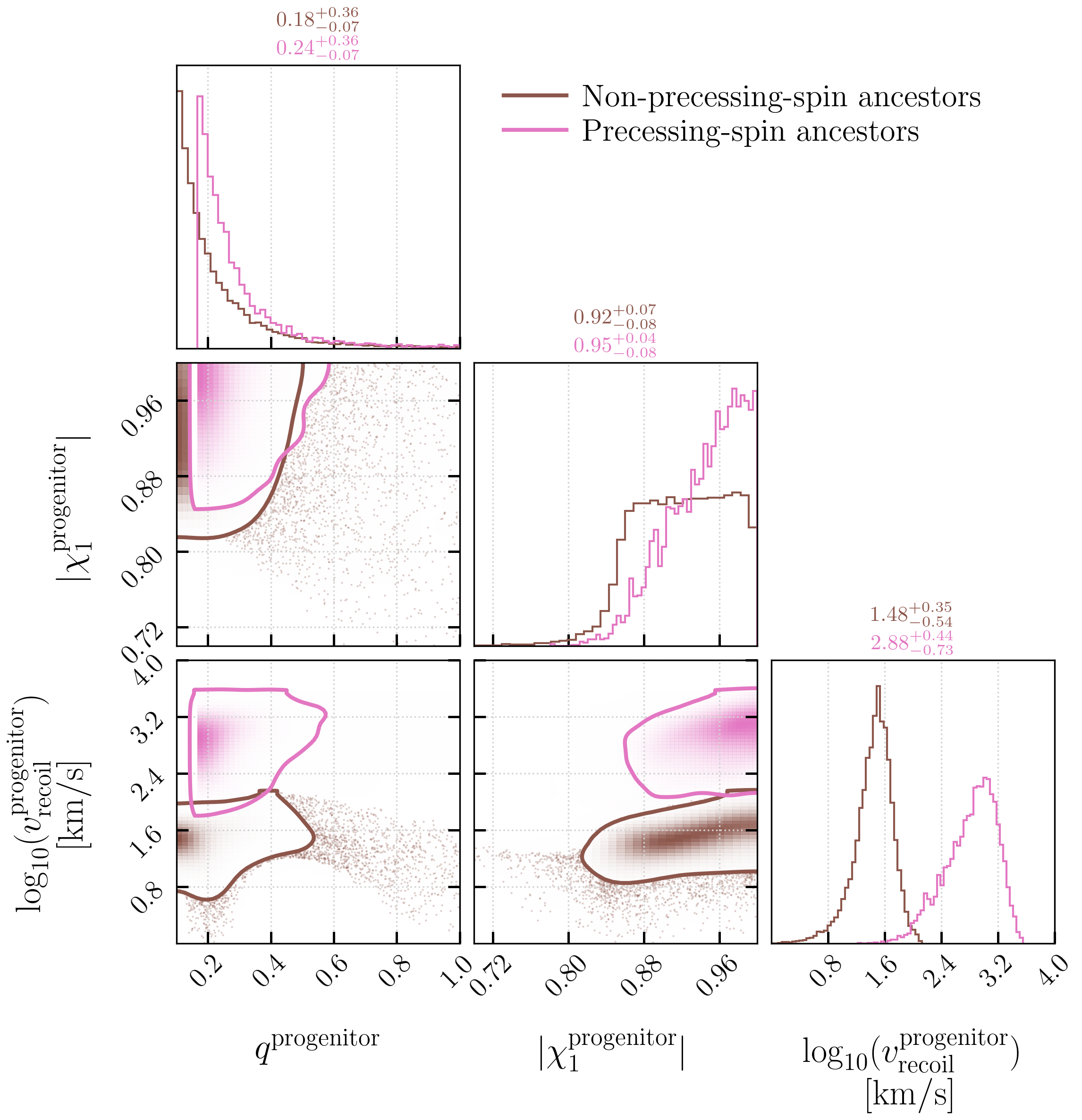}
    \caption{We show the inferred progenitor properties \emph{of the primary black hole} in the GW200114\_020818 binary under two scenarios: (i) a non-precessing ancestral binary (brown) and (ii) a precessing ancestral binary (pink). In the precessing case, there is a significant probability of large recoil velocities, implying that the remnant may not be retained in typical host environments. By contrast, in the non-precessing scenario, the recoil velocities are predominantly below $150\,\mathrm{km\,s^{-1}}$. More details are in Section~\ref{sec:formation_gw20}.}
    \label{fig:gw200114_m1_ancestral_posteriors}
\end{figure}

\begin{figure*}
    \centering
    \includegraphics[width=\textwidth]{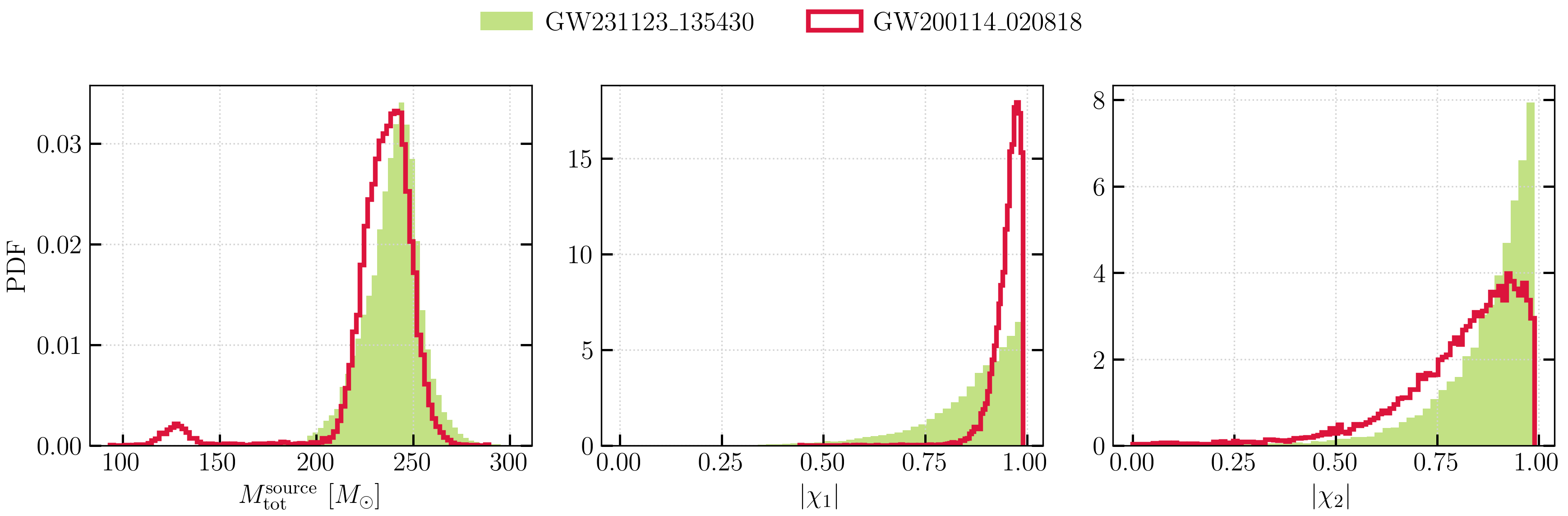}
    \caption{We show the parameters where GW200114\_020818 resembles GW231123\_135430. We compare the source-frame total mass $M_{\rm tot}^{\rm source}$ and the dimensionless spin magnitudes $|\chi_1|$ and $|\chi_2|$ as obtained from \texttt{NRSur7dq4} model. More details are in Section~\ref{sec:implications}.}
    \label{fig:gw231123_200114_comparison}
\end{figure*}

%%%%%%%%%%%%%%%%%%%%%%%%%%%%%%%%%%%%%%%%%%%%%%%%%%%%%%%%%%%%%%%%%%%%%%%%%%%%%%%%%%%
\subsection{Formation channel for GW200114\_020818}
\label{sec:formation_gw20}
%%%%%%%%%%%%%%%%%%%%%%%%%%%%%%%%%%%%%%%%%%%%%%%%%%%%%%%%%%%%%%%%%%%%%%%%%%%%%%%%%%%
These inferences have significant implications for the formation history of the constituent black holes in GW200114\_020818. The combination of a strongly negative effective inspiral spin, highly asymmetric mass ratio and massive nature of the black holes indicates that these black holes are unlikely to be formed through isolated binary evolution~\cite{Croon:2025gol}.
While the secondary mass, $m_{2}^{\rm source} = 34.9^{+4.0}_{-3.3}\,M_{\odot}$, is consistent with formation via direct stellar collapse, the inferred primary mass, $m_{1}^{\rm source} = 199.1^{+15.7}_{-16.9}\,M_{\odot}$, is too large to be explained by this channel.
It rather points toward a dynamical or hierarchical merger scenario, potentially occurring in globular clusters or nuclear star clusters, where spin orientations are expected to be largely isotropic and can naturally lead to negative values of $\chi_{\rm eff}$~\cite{Kalogera:1999tq,Gerosa:2018wbw}. In contrast, isolated binary evolution scenarios typically predict more aligned black hole spins as a result of tidal interactions and mass transfer~\cite{Kalogera:1999tq,Gerosa:2018wbw}.

A straightforward way to assess whether the inferred binary source properties of GW200114\_020818 are consistent with typical hierarchical merger scenarios in globular clusters is to compare them with publicly available few-body simulations of globular cluster evolution and BBH assembly and mergers. We utilize the \texttt{Cluster Monte Carlo} (\texttt{CMC}) simulations~\cite{Kremer:2019iul}~\footnote{\href{https://cmc.ciera.northwestern.edu/BBHmergers_CMCcatalog.dat}{https://cmc.ciera.northwestern.edu/BBHmergers\_CMCcatalog.dat}}, which span a grid of 128 globular cluster models with varying initial cluster masses, half-mass radii, metallicities, and galactocentric distances (Fig.~\ref{fig:cmc}). Note that the CMC initial conditions were roughly chosen to span the observed properties of Milky Way globular clusters~\cite{Kremer:2019iul}. We compare the source-frame masses $m_1^{\rm source}$ and $m_2^{\rm source}$ of all BBH mergers that occur within these simulated clusters to the inferred source masses of GW200114\_020818 obtained using different waveform models. We find that the source masses inferred for GW200114\_020818 ($199.1^{+15.7}_{-16.9}\,M_{\odot}$ and $34.9^{+4.0}_{-3.2}\,M_{\odot}$) are clear outliers relative to the typical outcomes of globular cluster evolution in the \texttt{CMC} simulations (Fig.~\ref{fig:cmc}; \textit{left panel}).
We then examine the spin magnitudes and reach a similar conclusion (Fig.~\ref{fig:cmc}; \textit{right panel}). 

Even though hierarchical mergers in \texttt{CMC} simulations are found to produce a small number of black holes with masses similar to those inferred for GW200114\_020818, achieving the high spin magnitudes ($|\chi_{1,2}| \ge 0.9$) is quite difficult. In hierarchical merger scenarios, most dynamically formed black holes have spin magnitudes below $0.8$. 
This suggests that alternative formation channels for GW200114\_020818 should be explored.
Possible formation pathways for this binary include formation within an active galactic nucleus (AGN) environment and Population~III remnants that subsequently grew via coherent, misaligned gas accretion~\cite{Bartos:2025pkv}. An additional viable channel is chemically homogeneous evolution, involving progenitors with masses exceeding the PISN gap~\cite{2010AIPC.1314..291D,Popa:2025dpz}.

However, it must be acknowledged that, although typical astrophysical formation scenarios may struggle to explain binaries such as GW200114\_020818, within general relativity (GR), the formation of black holes with near-extremal spins, $|\chi_{1,2}| \gtrsim 0.9$, is not forbidden and can naturally arise through hierarchical mergers.
Motivated by this, we consider the 90\% credible interval of the inferred primary spin magnitude and investigate the properties of a possible hierarchical progenitor. Assuming that the primary black hole formed through a previous merger, we use \texttt{NRSur7dq4Remnant} to infer the spins of the ancestral binary. For a precessing ancestral system, we find that the spin of the primary progenitor must be $0.95^{+0.04}_{-0.05}$, while the secondary progenitor spin remains largely unconstrained (Fig.~\ref{fig:gw200114_m1_ancestral_posteriors}). In this scenario, the associated recoil kick velocity is $768^{+743}_{-491}\,\mathrm{km\,s^{-1}}$, implying that many such merger remnants would not be retained in typical globular clusters or dwarf galaxies, though retention would remain possible in AGNs. This implies that, if the primary black hole in GW200114\_020818 formed through a hierarchical merger, the most plausible host environment would be an AGN. Note that we have restricted \texttt{NRSur7dq4Remnant} to its validation domain with NR simulations $q\geq 0.167$. Extending its range to lower $q$ can lower our recoil kick velocity estimates.

By contrast, if the ancestral binary were non-precessing, using \texttt{NRSur3dq8Remnant}, the inferred primary progenitor spin would be $0.92^{+0.06}_{-0.06}$ (while the secondary progenitor spin remains unconstrained), with a substantially smaller recoil velocity of $30^{+19}_{-14}\,\mathrm{km\,s^{-1}}$ (Fig.~\ref{fig:gw200114_m1_ancestral_posteriors}). Such remnants would be readily retained in globular clusters. However, aligned-spin mergers are generally disfavored in dynamically assembled environments like globular clusters and are more likely to be associated with AGNs. 
In both cases, we find a preference for asymmetric ancestral mass ratios.

%%%%%%%%%%%%%%%%%%%%%%%%%%%%%%%%%%%%%%%%%%%%%%%%%%%%%%%%%%%%%%%%%%%%%%%%%%%%%%%%%%%
\subsection{Similarity between GW200114\_020818 and GW231123\_135430}
\label{sec:gw231123}
%%%%%%%%%%%%%%%%%%%%%%%%%%%%%%%%%%%%%%%%%%%%%%%%%%%%%%%%%%%%%%%%%%%%%%%%%%%%%%%%%%%
Next, we explicitly highlight the similarities between GW200114\_020818 and GW231123\_135430. In particular, we focus on the source-frame total mass $M_{\rm tot}^{\rm source}$ and the dimensionless spin magnitudes $|\chi_1|$ and $|\chi_2|$ as obtained from the \texttt{NRSur7dq4} model in Fig.~\ref{fig:gw231123_200114_comparison}. We find that the source-frame total masses of the two events are very similar, as are the spin magnitudes of the secondary black holes. However, GW200114\_020818 exhibits even larger spin magnitude for the primary black hole compared to GW231123\_135430. In addition, while GW231123\_135430 is a nearly equal-mass binary, GW200114\_020818 is highly asymmetric, with mass ratios $q \lesssim 0.25$.

\begin{figure}
    \centering
    \includegraphics[width=\columnwidth]{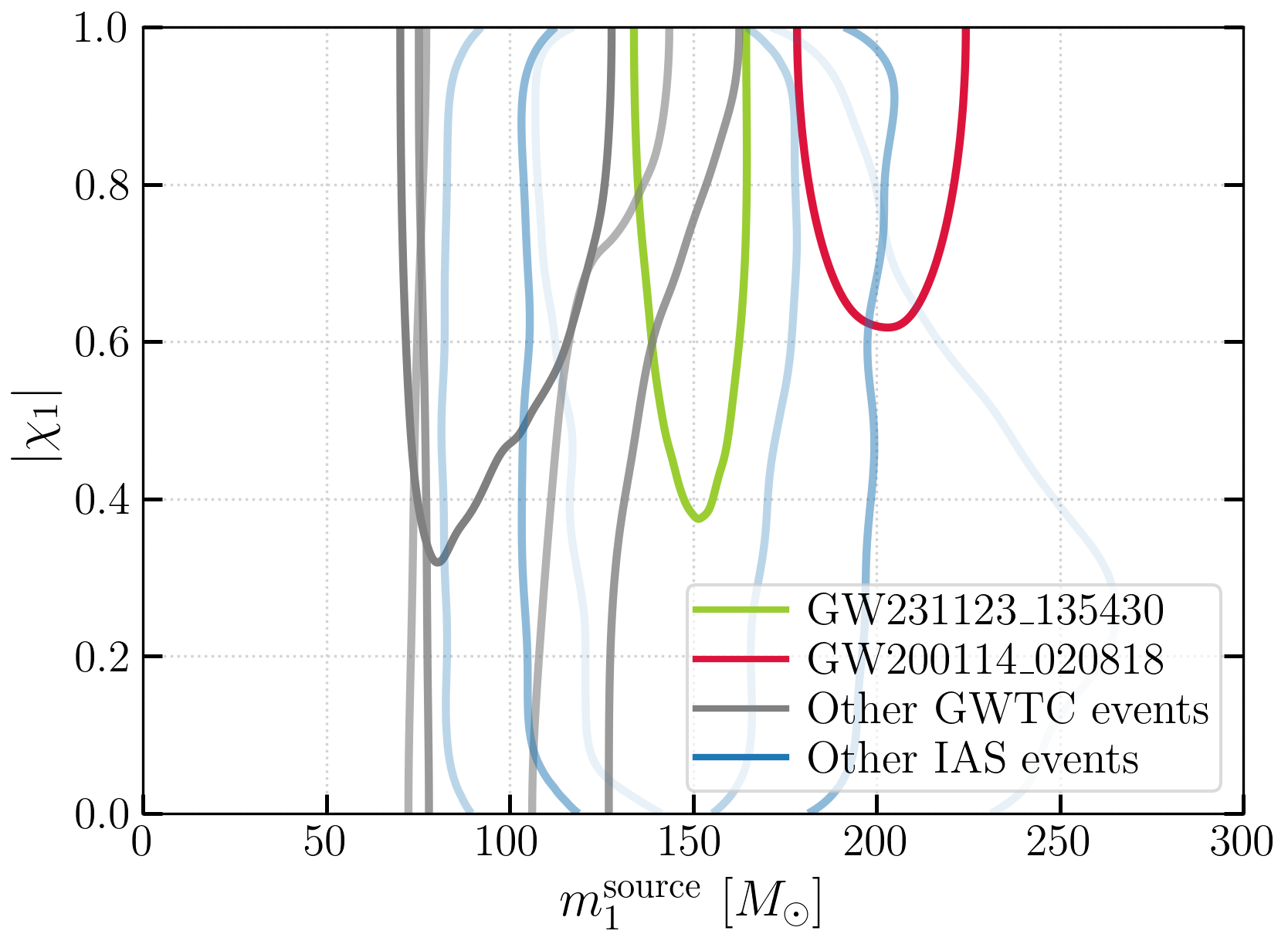}
    \caption{We show the inferred primary masses and spin magnitudes for events with large values of both quantities, drawn from the GWTC and IAS catalogs. For comparison, we also include the inferred parameters for GW200114\_020818 (crimson) and GW231123\_135430 (yellow-green). These events could either hint at rare outliers from the currently-known population, or they could hint at a new massive, highly spinning sub-population of BHs. More details are in Section~\ref{sec:gw231123}.}
    \label{fig:massive_spinning_pop}
\end{figure}

%%%%%%%%%%%%%%%%%%%%%%%%%%%%%%%%%%%%%%%%%%%%%%%%%%%%%%%%%%%%%%%%%%%%%%%%%%%%%%%%%%%
\subsection{Emergence of massive, highly spinning population}
\label{sec:newpop}
%%%%%%%%%%%%%%%%%%%%%%%%%%%%%%%%%%%%%%%%%%%%%%%%%%%%%%%%%%%%%%%%%%%%%%%%%%%%%%%%%%%
The unusual source properties of GW200114\_020818 and GW231123\_135430, particularly their large masses and high spin magnitudes, raise an important question: are these events rare outliers, or do they represent the emergence of a previously unrecognized population of BBHs? Such a population may originate from a distinct formation channel that is not yet fully understood. To illustrate this potential population, Fig.~\ref{fig:massive_spinning_pop} shows the inferred primary source masses and primary spin magnitudes events characterized by large masses and spins. We draw the events both from the GWTC and IAS catalogs. 
For comparison, we also include the inferred parameters of GW200114\_020818 and GW231123\_135430.
We can immediately see that there are around three events in the GWTC catalog for which the primary mass has some support beyond $100\,M_{\odot}$. On the other hand, in the IAS catalog, we identify three additional events that populate the same mass range as GW200114\_020818 and GW231123\_135430. 
We, however, find that most of these events have poorly constrained spin magnitudes due to their low SNR.
Taken together, these events hint at the possible emergence of a sub-population of massive, highly spinning black holes.
As more binaries are detected in the coming years, it will become possible to address this question more definitively. In the meantime, it is essential to determine whether such a population can be explained within the framework of known astrophysical formation channels.

%%%%%%%%%%%%%%%%%%%%%%%%%%%%%%%%%%%%%%%%%%%%%%%%%%%%%%%%%%%%%%%%%%%%%%%%%%%%%%%%%%%
%%%%%%%%%%%%%%%%%%%%%%%%%%%%%%%%%%%%%%%%%%%%%%%%%%%%%%%%%%%%%%%%%%%%%%%%%%%%%%%%%%%
\section{Waveform Modeling Implications}
\label{sec:modellingimplications}
%%%%%%%%%%%%%%%%%%%%%%%%%%%%%%%%%%%%%%%%%%%%%%%%%%%%%%%%%%%%%%%%%%%%%%%%%%%%%%%%%%%
%%%%%%%%%%%%%%%%%%%%%%%%%%%%%%%%%%%%%%%%%%%%%%%%%%%%%%%%%%%%%%%%%%%%%%%%%%%%%%%%%%%
Finally, we discuss the waveform-modeling implications of GW200114\_020818 and GW231123\_135430. 
The exceptional nature of these two events not only makes them astrophysically puzzling, but also  tests the current limits of waveform-modeling efforts. In particular, none of the waveform models currently available are known to be reliably accurate in the highly spinning regime (which appears to be the case for both GW200114\_020818 and GW231123\_135430).  there are almost no SXS NR simulations of precessing binaries with spin magnitudes $|\chi_{1,2}| \gtrsim 0.8$. We illustrate this in Fig.~\ref{fig:combined_posteriors}, where we show the 
\texttt{NRSur7dq4} posteriors for these two events together with the parameter values of the available SXS NR simulations (along with simulations from the RIT and MAYA catalogs). 

A similar limitation applies to 
highly asymmetric binaries with mass ratios $q \lesssim 0.25$ (relevant for GW200114\_020818). This is largely because only a handful of precessing NR simulations exist for mass ratios below $0.25$. 
These sporadic coverage of the highly spinning, highly asymmetric mass parameter space leads to significant waveform systematics in GW200114\_020818 and GW231123\_135430~\cite{LIGOScientific:2025rsn}.
These events therefore provide strong motivation to extend the coverage of NR simulations and, in turn, improve the validity range of waveform models.

%%%%%%%%%%%%%%%%%%%%%%%%%%%%%%%%%%%%%%%%%%%%%%%%%%%%%%%%%%%%%%%%%%%%%%%%%%%%%%%%%%%
%%%%%%%%%%%%%%%%%%%%%%%%%%%%%%%%%%%%%%%%%%%%%%%%%%%%%%%%%%%%%%%%%%%%%%%%%%%%%%%%%%%
\section{Conclusions}
\label{sec:discussions}
%%%%%%%%%%%%%%%%%%%%%%%%%%%%%%%%%%%%%%%%%%%%%%%%%%%%%%%%%%%%%%%%%%%%%%%%%%%%%%%%%%%
%%%%%%%%%%%%%%%%%%%%%%%%%%%%%%%%%%%%%%%%%%%%%%%%%%%%%%%%%%%%%%%%%%%%%%%%%%%%%%%%%%%
In this work, we have reanalyzed GW190711\_030756 and GW200114\_020818, the two loudest candidates in the IAS catalog which were not in GWTC-3. We used the NR surrogate waveform model \texttt{NRSur7dq4}. Both systems favor asymmetric mass ratios and unusually large total masses, placing them among the most extreme BBH mergers studied so far.
While GW190711\_030756 has $p_{\rm astro}=0.99$ and an SNR of $\sim 9.5$, GW200114\_020818 has $p_{\rm astro}=0.71$ and an SNR of $\sim 13$.

Both events stand out due to their highly asymmetric mass configurations, with inferred mass ratios of $\sim 0.35$ and $\sim 0.17$ for GW190711\_030756 and GW200114\_020818, respectively. For most other parameters, GW190711\_030756 does not exhibit any notable exceptional features. 
By contrast, GW200114\_020818 presents a markedly different picture. This event is distinguished by its extreme spin properties and its similarity to GW231123\_135430. It is inferred to be a highly massive BBH merger in which at least one component black hole is rapidly spinning, with \texttt{NRSur7dq4} favoring near-extremal spins. The system shows strong evidence for anti-aligned spins, resulting in the most negative effective inspiral spin measured to date. Such a configuration is difficult to reconcile with isolated binary evolution, which generally predicts spin--orbit alignment, and instead favors a dynamical formation channel where spin orientations are expected to be nearly isotropic. However, it is also quite difficult to form BHs with almost near extremal spin in globular clusters. Therefore, the likely origin could be in AGNs or some other astrophysical mechanism must be invoked.
Although waveform systematics influence the precise parameter estimates, the qualitative conclusions remain robust: GW200114\_020818 is a strongly asymmetric, highly massive system characterized by pronounced spin effects. The inferred remnant mass lies firmly in the IMBH regime, and the predicted recoil velocity implies a negligible retention probability in globular clusters but efficient retention in environments with higher escape velocities, such as nuclear star clusters and elliptical galaxies.

Furthermore, GW200114\_020818 closely resembles GW231123\_135430 in terms of both the total mass and the spin magnitudes. Taken together, GW200114\_020818, particularly when considered alongside GW231123\_135430, provides suggestive evidence for the emergence of a population of highly spinning, massive black holes possibly via dynamical assembly and potentially hierarchical black hole growth in high-escape-velocity environments. Future improvements in waveform modeling at smaller mass ratios, along with additional detections of high-mass BBH mergers, will be crucial for determining whether such systems represent rare outliers or the early members of an emerging population (Fig.~\ref{fig:massive_spinning_pop}).

Finally, we inspect the spin-angle dynamics of these two events and find no evidence for exotic behaviors such as up--down instability, transitional precession, or spin--orbit resonances.
We have made our results publicly available at \href{https://github.com/tousifislam/GW190711\_GW200114}{https://github.com/tousifislam/GW190711\_GW200114}.

%%%%%%%%%%%%%%%%%%%%%%%%%%%%%%%%%%%%%%%%%%%%%%%%%%%%%%%%%%%%%%%%%%%%%%%%%%%%%%%%%%%%%%%%%%%%%%%%%%%%%%%%
%%%%%%%%%%%%%%%%%%%%%%%%%%%%%%%%%%%%%%%%%%%%%%%%%%%%%%%%%%%%%%%%%%%%%%%%%%%%%%%%%%%%%%%%%%%%%%%%%%%%%%%%
\begin{acknowledgments}
T.I. is supported in part by the National Science Foundation under Grant No. NSF PHY-2309135 and the Gordon and Betty Moore Foundation Grant No. GBMF7392. 
Use was made of computational facilities purchased with funds from the National Science Foundation (CNS-1725797) and administered by the Center for Scientific Computing (CSC). The CSC is supported by the California NanoSystems Institute and the Materials Research Science and Engineering Center (MRSEC; NSF DMR 2308708) at UC Santa Barbara. 
D.W. is supported by NSF Grants No.~AST-2307146, No.~PHY-2513337, No.~PHY-090003, and No.~PHY-20043, by NASA Grant No.~21-ATP21-0010, by John Templeton Foundation Grant No.~62840, by the Simons Foundation [MPS-SIP-00001698, E.B.], by the Simons Foundation International [SFI-MPS-BH-00012593-02], and by Italian Ministry of Foreign Affairs and International Cooperation Grant No.~PGR01167.

This research has made use of data, software and/or web tools obtained from Zenodo and the Gravitational Wave Open Science Center (\url{https://www.gw-openscience.org/}), a service of LIGO Laboratory, the LIGO Scientific Collaboration and the Virgo Collaboration. LIGO Laboratory and Advanced LIGO are funded by the United States National Science Foundation (NSF) as well as the Science and Technology Facilities Council (STFC) of the United Kingdom, the Max-Planck-Society (MPS), and the State of Niedersachsen/Germany for support of the construction of Advanced LIGO and construction and operation of the GEO600 detector. Additional support for Advanced LIGO was provided by the Australian Research Council. Virgo is funded, through the European Gravitational Observatory (EGO), by the French Centre National de Recherche Scientifique (CNRS), the Italian Istituto Nazionale di Fisica Nucleare (INFN) and the Dutch Nikhef, with contributions by institutions from Belgium, Germany, Greece, Hungary, Ireland, Japan, Monaco, Poland, Portugal, Spain.
\end{acknowledgments}
%%%%%%%%%%%%%%%%%%%%%%%%%%%%%%%%%%%%%%%%%%%%%%%%%%%%%%%%%%%%%%%%%%%%%%%%%%%%%%%%%%%%%%%%%%%%%%%%%%%%%%%%
%%%%%%%%%%%%%%%%%%%%%%%%%%%%%%%%%%%%%%%%%%%%%%%%%%%%%%%%%%%%%%%%%%%%%%%%%%%%%%%%%%%%%%%%%%%%%%%%%%%%%%%%

%%%%%%%%%%%%%%%%%%%%%%%%%%%%%%%%%%%%%%%%%%%%%%%%%%%%%%%%%%%%%%%%%%%%%%%%%%%%%%%%%%%%%%%%%%%%%%%%%%%%%%%%
\bibliography{references}
%%%%%%%%%%%%%%%%%%%%%%%%%%%%%%%%%%%%%%%%%%%%%%%%%%%%%%%%%%%%%%%%%%%%%%%%%%%%%%%%%%%%%%%%%%%%%%%%%%%%%%%%

\appendix

\begin{figure}
    \centering
    \includegraphics[width=\columnwidth]{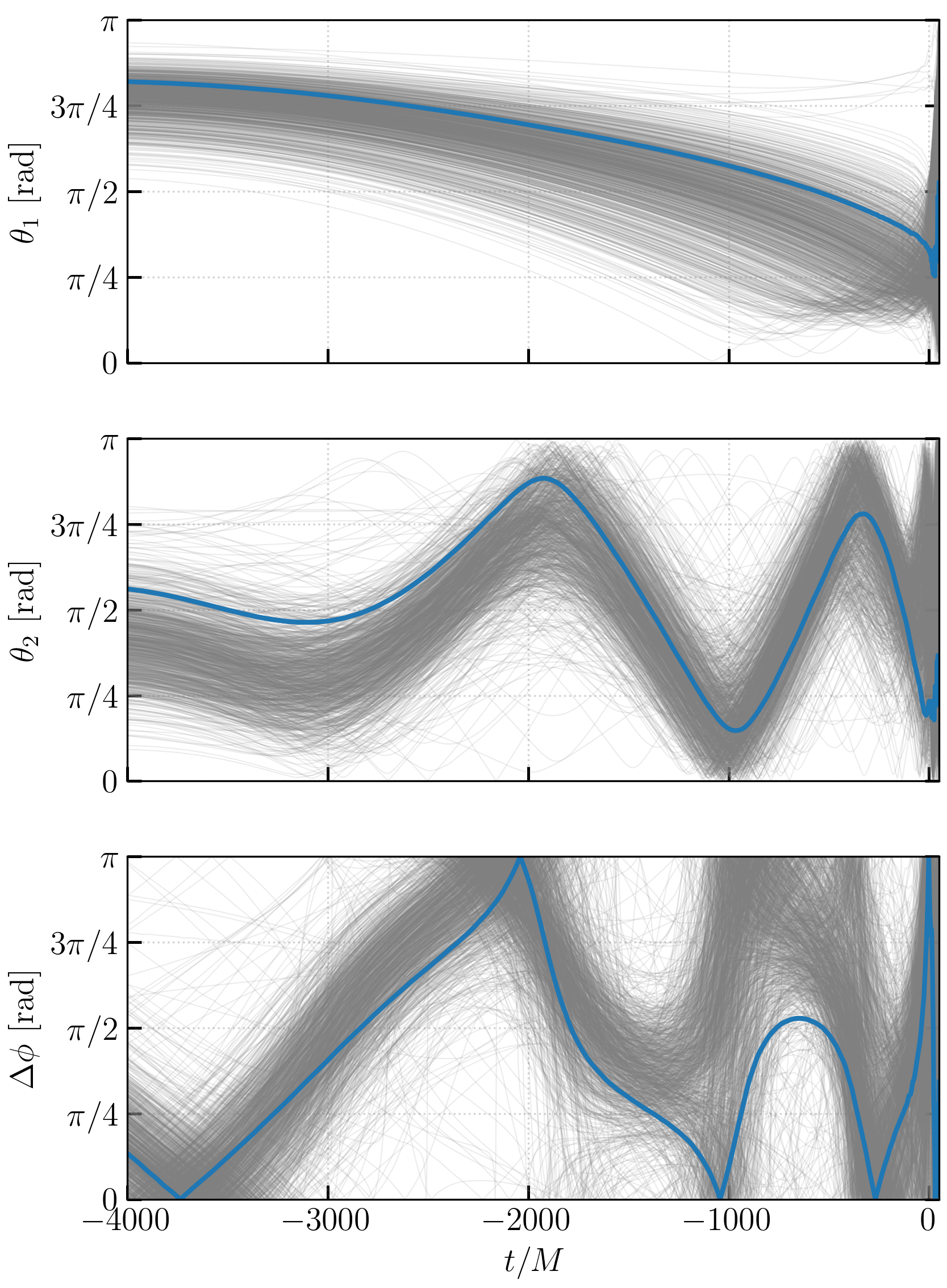}
    \caption{We show the time evolution of the tilt angles $\theta_1$ and $\theta_2$, as well as $\Delta\phi$, the angle between the in-plane spin angles $\phi_1$ and $\phi_2$, for 1000 randomly selected posterior samples of GW200114\_020818 (gray). For comparison, we also highlight the spin evolution corresponding to the maximum-likelihood estimate (dark blue). More details are in Appendix~\ref{sec:dynamics}
    }
    \label{fig:dynamics}
\end{figure}

%%%%%%%%%%%%%%%%%%%%%%%%%%%%%%%%%%%%%%%%%%%%%%%%%%%%%%%%%%%%%%%%%%%%%%%%%%%%%%%%%%%
%%%%%%%%%%%%%%%%%%%%%%%%%%%%%%%%%%%%%%%%%%%%%%%%%%%%%%%%%%%%%%%%%%%%%%%%%%%%%%%%%%%
\section{Inspecting the binary dynamics}
\label{sec:dynamics}
%%%%%%%%%%%%%%%%%%%%%%%%%%%%%%%%%%%%%%%%%%%%%%%%%%%%%%%%%%%%%%%%%%%%%%%%%%%%%%%%%%%
%%%%%%%%%%%%%%%%%%%%%%%%%%%%%%%%%%%%%%%%%%%%%%%%%%%%%%%%%%%%%%%%%%%%%%%%%%%%%%%%%%%
The final component of our investigation concerns the binary dynamics of both events, with particular emphasis on the evolution of the spin angles $\theta_1$, $\theta_2$, $\phi_1$, and $\phi_2$. For the in-plane spins, rather than examining the individual azimuthal angles, we study their relative separation $\Delta\phi$. We use \texttt{NRSur7dq4} to evaluate the spin evolution as a function of geometric time (where we set $G=c=1$).
For GW190711\_030756, the spin angles are largely unconstrained, and consequently their evolution is not informative. Therefore, we do not show the spin angles evolution. In contrast, the spin angles for GW200114\_020818 are well constrained. We find that $\theta_1$, $\theta_2$, and $\Delta\phi$ remain well constrained throughout the inspiral, with sudden changes near the end of the evolution. The corresponding time evolution is shown in Fig.~\ref{fig:dynamics}.

\begin{figure}
    \centering
    \includegraphics[width=\columnwidth]{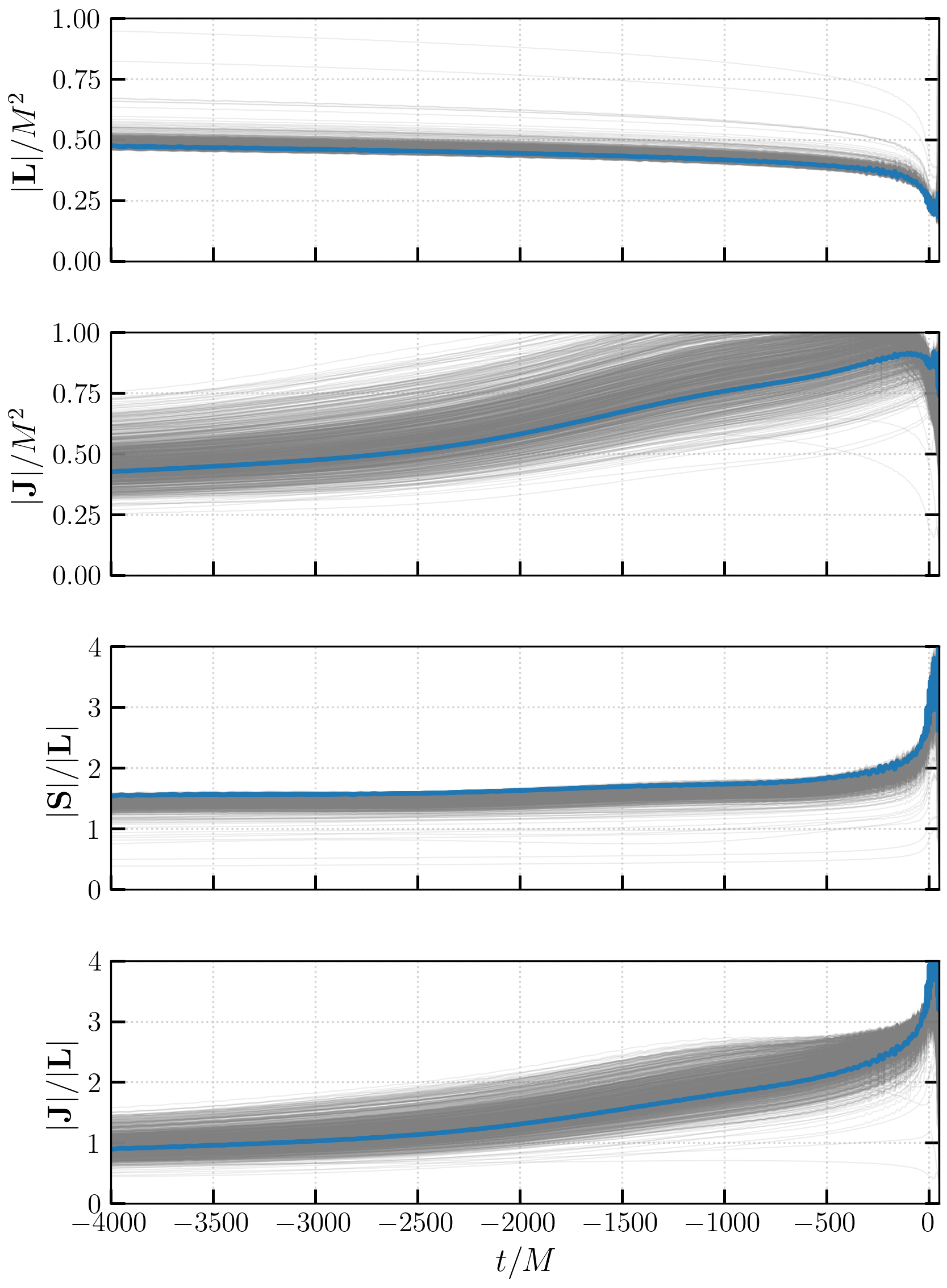}
    \caption{We show the time evolution of the total angular momentum $\mathbf{J}$, orbital angular momentum $\mathbf{L}$, and total spin $\mathbf{S}$, together with their relative magnitudes and ratios, for 1000 randomly selected posterior samples of GW200114\_020818 (gray). For comparison, we also highlight the spin evolution corresponding to the maximum-likelihood estimate (dark blue). More details are in Appendix~\ref{sec:dynamics}.}
    \label{fig:gw200114_angular_momentum}
\end{figure}

We use the inferred binary dynamics to investigate three possible signatures of nontrivial spin behavior: (i) up--down instability~\cite{Gerosa:2015hba,Lousto:2016nlp,Johnson-McDaniel:2021rvv,Varma:2020bon}, (ii) transitional precession~\cite{PhysRevD.49.6274}, and (iii) spin--orbit resonance~\cite{Schnittman:2004vq}.
In the up--down instability scenario, the binary begins with $\vec{S_1} \parallel +\vec{L}$ and $\vec{S_1} \parallel -\vec{L}$, with small perturbations. Due to the instability, these perturbations grow over time, leading to strong precession near merger. In this case, the tilt angles evolve toward a common value close to merger,
\begin{equation}
\cos\theta_{1,2} \rightarrow 
\frac{|\chi_1| - q |\chi_2|}{|\chi_1| + q |\chi_2|},
\end{equation}
while the difference in the in-plane spin azimuthal angles approaches zero: $\Delta\phi \rightarrow 0$.
In transitional precession, the total angular momentum $\mathbf{J}$ temporarily approaches zero when the orbital angular momentum ($\mathbf{L}$) nearly cancels the total spin ($\mathbf{S}$),
\begin{equation}
\mathbf{L} \approx -\mathbf{S},
\end{equation}
causing $\mathbf{J}$ to pass through a minimum magnitude and resulting in characteristic changes in the precession dynamics.
Finally, spin--orbit resonances are characterized by approximate coplanarity and libration of the in-plane spin azimuthal separation, with
\begin{equation}
\Delta\phi \approx 0 \quad \text{or} \quad \pi,
\end{equation}
rather than circulation over the full angular range.
Since GW200114\_020818 exhibits evidence for both large spins and significant precession, it is natural to examine whether any of these dynamical features are present in this system. In Fig.~\ref{fig:gw200114_angular_momentum}, we show these quantities as they evolve with time.

We find no evidence for any of the three dynamical features in the inferred spin-angle evolution. In particular, $\theta_1$ and $\theta_2$ do not converge to a common value during the evolution. Furthermore, $\Delta\phi$ does not approach zero near merger.

\end{document}